# Revisit to the Inverse Exponential Radon Transform[1]

Jiangsheng (Jason) You

Cubic Imaging LLC, 18 Windemere Dr., Andover, MA 01810

jshyou@gmail.com

## Abstract

This revisit gives a survey on the analytical methods for the inverse exponential Radon transform which has been investigated in the past three decades from both mathematical interests and medical applications such as nuclear medicine emission imaging. The derivation of the classical inversion formula is through the recent argument developed for the inverse attenuated Radon transform. That derivation allows the exponential parameter to be a complex constant, which is useful to other applications such as magnetic resonance imaging and tensor field imaging. The survey also includes the new technique of using the finite Hilbert transform to handle the exact reconstruction from 180º data. Special treatment has been paid on two practically important subjects. One is the exact reconstruction from partial measurements such as half-scan and truncated-scan data, and the other is the reconstruction from diverging-beam data. The noise propagation in the reconstruction is touched upon with more heuristic discussions than mathematical inference. The numerical realizations of several classical reconstruction algorithms are included. In the conclusion, several topics are discussed for more investigations in the future.

## Table of content

---

[1] This review was written in 2007. The comments in this review only reflect the author's personal opinion. Recently the author revisited previously unfinished works and intended to publish some results that still look new.

# I. Introduction

First we discuss some basics on the single photon emission computed tomography (SPECT), which is one of most important applications of the exponential Radon transform (ERT). The clinic study by SPECT includes two steps: 1) inject radionuclide tracer (e.g., $Tc^{99}$, $Tl^{201}$) into patient's body for blood to carry the tracer to specific tissues, and 2) use scintillation camera to detect the emitted γ-photons from outside of the body. This procedure can measure the blood that flows to different body tissues through estimating the distribution of the radionuclide tracer inside the body. Detected photons on the scintillation detector behind each collimation hole come from a narrow strip (illustrated as thick red line in Fig 1) and have been attenuated by the body tissue. The instrument physics and mechanics of the scintillation camera for photon counts detection were detailed in the pioneer work [1]. The integration of tomographic technique with the scintillation camera started in those early works [2-6]. Readers should keep in mind that the detected photon counts from each collimation hole are proportional to the accumulated radionuclide tracer along the red line in Fig 1. Restricted on a planar section as shown in Fig 1, we use $f(x, y)$ to represent the density distribution of radionuclide tracer at point $P$ on that section. The goal of SPECT imaging is to estimate $f(x, y)$ from the detected photon counts. In medical imaging, this process is usually called the image reconstruction from the line integrals. The SPECT data formation shall be adopted in this paper although there are many other applications of the ERT [7-9].

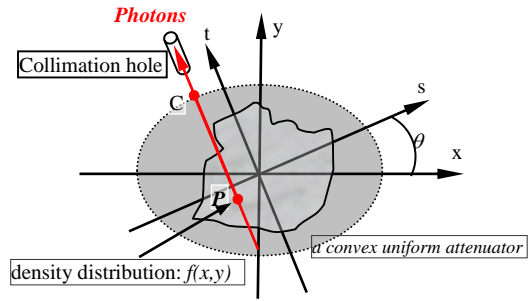

Fig 1: Basics of the photon detection in SPECT and the data formation under parallel-beam scanning geometry. The attenuation is uniform inside the ellipse. The irregular-shaped area inside the ellipse represents the region in which radionuclide tracer is distributed.

Next we introduce several mathematic notions. Denote by $R^2$ the two-dimensional (2D) plane space with point representation by $\vec{r} = (x, y)$ in Cartesian coordinates and $\vec{r} = (r, \varphi)$ in polar coordinates, respectively. Define vectors $\vec{\theta} = (\cos\theta, \sin\theta)$ and $\vec{\theta}^\perp = (-\sin\theta, \cos\theta)$, here $\vec{\bullet}^\perp$ means 90° counterclockwise rotation of a vector. Take $I = [-1, 1]$, and let $D = \{\vec{r} \in R^2 : |\vec{r}| = \sqrt{x^2 + y^2} < 1\}$ and $S^1$ be the unit disk and circle of $R^2$, respectively. By scaling, we may assume $f(x, y) \equiv 0$ outside of $D$, and the linear attenuation in SPECT is non-negative, i.e., $\mu \geq 0$, inside the ellipse in Fig 1. For example, all these assumptions can be met in brain SPECT after removing the attenuation of the skull. Let $L(\theta, s)$ be the distance between the point $C$ and $s$-axis in Fig. 1. Then the photon counts passing through the collimation hole at $(\theta, s)$ is proportional to $\int_{PC} f(P + t\vec{\theta}^\perp) e^{-\mu(L(\theta,s)-t)} dt$, here the integral is taken along the line $PC$. Since the weight $e^{-\mu L(\theta,s)}$ can be precomputed. After scaling, the modified photon counts can be expressed as the ERT which is defined by

$$[\mathbf{R}_\mu f](\theta, s) = \int_{-\infty}^{\infty} f(s\vec{\theta} + t\vec{\theta}^\perp) e^{\mu t} dt . \tag{1.1}$$

For simplicity, we use $p(\theta, s)$ for $[\mathbf{R}_\mu f](\theta, s)$. In SPECT, both $p(\theta, s)$ and $f(x, y)$ take non-negative real values. However, in some applications described in [7-9], $f(x, y)$ and $\mu$ may take imaginary values. Throughout this paper, we assume that $f(x, y)$ is a continuously smooth complex-valued function of compact support in $D$ and $\mu$ is a complex constant. $\mathbf{R}_\mu$ is a mapping from functions on $D$ to functions of $S^1 \times I$. For a function $g(\theta, s)$ of $S^1 \times I$, the adjoint $\mathbf{R}_\mu^*$ (usually called as the back-projection operation) is defined by

$$[\mathbf{R}_\mu^* g](x, y) = \int_0^{2\pi} e^{-\mu \vec{r} \cdot \vec{\theta}^\perp} g(\theta, \vec{r} \cdot \vec{\theta}) d\theta . \qquad (1.2)$$

The common task in these applications is to find $f(x, y)$ from the (modified) measurement $p(\theta, s)$ by using various physical devices. We call such procedure as the image reconstruction from projections. When the attenuation is not present, i.e., $\mu = 0$, this issue was originally introduced in the historic works [10, 11] long time ago. Since then, there have been tremendous progresses being made on both theoretical studies on the inverse ERT (IERT) and engineering developments of using the IERT in last three decades. The most theoretically beautiful and numerically efficient image reconstruction method is the filtered back-projection (FBP) algorithm which was first developed for computed tomography (CT) image reconstruction in [12]. The extension of FBP algorithm to the IERT was first obtained in [13] while other types of algorithms were investigated at the same time. In this paper, I give a survey on the available analytical methods for the IERT. Meanwhile, I always feel deeply grateful to all contributions in the field even many of them have not been mentioned in the paper. In particular, very few of iterative methods are discussed in this paper although these methods have played a very important role along the history.

The goal of the IERT is to estimate $f(x, y)$ from the exponentially-weighted projection $p(\theta, s)$ defined by (1.1). When $\mu = 0$, it is easy to observe the symmetry of $p(\theta, -s) = p(\theta + \pi, s)$ and the Fourier slice theorem [12], thus the derivation of FBP algorithm is rather straightforward. However, the IERT does not appear to be obvious when $\mu \neq 0$ as revealed in the history. For the case of $\mu \in R^1$, a lot of efforts have been pursued in last three decades on both theoretical investigations and numerical studies on the IERT and its applications. Those works [13-28] seem to be quite representative of the analytical methods for the IERT. Undoubtedly, work [13] provided the most fundamental contributions to the explicit formula for the IERT. Other theoretical extension of the FBP-type inversion was achieved in [14-16] for the angular-dependent attenuation, but without being used in practical applications. At the same time, many results similar to Theorem 1 of [13] were investigated in works such as [17-23] from SPECT considerations. The Fourier method in these works, by and large, can be simply described as the frequency-domain expression of Theorem 1 of [13] by using the Fourier transform (FT) and Fourier series expansion (FSE), which was clearly described in Theorem 2 of [13]. Another interesting formula was derived in [24] using the integral geometry theory. Similar to the attenuation-free case, the FBP inversion formula of [13] also can be expressed in the circular harmonic expansion (CHE) [25-26]. This method can be regarded as an extension of Cormack's famous work [11] to the IERT. Also, the procedure of back-projection followed by a 2D filtering was investigated in [27-29]. Due to the breakthrough from [30] on the inverse attenuated Radon transform and many subsequent developments in [31-35], it has been found that the IERT can be treated with these new techniques used for the inverse attenuated Radon transform. In this review, we will give the derivation of the classical FBP inversion formula using the argument from [31, 35].

Keep in mind that all above mentioned works need the projection data to be available over $\theta \in [0, 2\pi)$. On the other hand, the scanning range over $[0, \pi]$ is sufficient enough for an exact reconstruction due to the strong symmetry of the attenuation-free projection. With the increasing interests in the exact reconstruction from partial projection data, many works have been done to deal with the exact reconstruction from partially attenuated data in [36-47]. These works mainly fall into two categories. One is using the FBP procedure and the other is reducing the reconstruction to the inverse Hilbert transform. In author's view, the method suggested in [38] is quite creative to handle the IERT with 180° data. The method in [38] consists of derivative, back-projection and inverse weighted Hilbert transform (DBH). Compared with the FBP-type inversion, the DBH has those advantages: 1) intrinsic of using the data over $[0, \pi]$ and 2) adaptive to both parallel- and fan-beam data. More details about the DBH method will be explored in this paper. In practical applications, the fan- and cone-bin data formations are very important. Those works [24-25, 29, 48-55] presented some efficient methods. The detailed geometry settings for the fan- and cone-bin data will be given in this review. Noise analysis in the image reconstruction is a frequently discussed topic, but most time lacks a solid mathematical rationale except many heuristic observations and inconclusive numerical results. The early works [56, 57] indicated that the FBP algorithm should be optimal under certain criterion. In [22], it was claimed that certain weighted Fourier method could be nearly optimal under different criterion. Recently, a new relation for the Poisson data was derived in [58] to design numerical filters. Intuitively, mixing the statistic rule with a completely determinant analytical process may be difficult to be justified with solid mathematical rationale. One gap pointed out in [59] is that reducing the variance of the projection data does not necessarily mitigate the noise in the reconstructed image when the range condition is involved. A filter may only improve the visual perception if it does not reduce the noise in the reconstructed image. The recent work [60] does not address this gap either.

The paper is presented in the following order. Section II includes a preliminary on the mathematics used in the derivation of many inversion formulas. Section III is the main component in this review and provides many inversion formulas and the derivations. Section IV addresses the exact reconstruction from partial data and Section V focuses on the fan- and cone-bin data formation. Section VI presents numerical simulation results for several classical reconstruction algorithms. Finally, Section VII concludes the review with some interesting topics worth more investigations in the future.

## II. A Mathematical Preliminary

In this section, I collect some frequently used mathematical results in the literature for the study of IERT. Let $\tilde{f}(\xi_x, \xi_y)$ be the 2D FT of $f(x, y)$ and $\tilde{p}(\theta, \omega)$ denote the one-dimensional (1D) FT of $p(\theta, l)$ with respect to the second variable. Throughout this paper, $\text{sgn}(\cdot)$ is the sign function and $\delta(\vec{r})$ represents the Dirac delta function of $R^2$. The 1D FT and its inverse take the following definitions:

$$\tilde{g}(\omega) = \frac{1}{\sqrt{2\pi}} \int_{-\infty}^{\infty} g(s) e^{-is\omega} ds, \quad g(s) = \frac{1}{\sqrt{2\pi}} \int_{-\infty}^{\infty} \tilde{g}(\omega) e^{is\omega} d\omega. \tag{2.1}$$

The choice of definition (2.1) is to keep the symmetry for both the FT and its inverse. By (2.1), the FTs of functions $\cos(\mu s)$ with real $\mu$ and $1/s$ in the distribution sense can be expressed as

$$\frac{\sqrt{2\pi}}{2}[\delta(\omega + \mu) + \delta(\omega - \mu)], \text{ and } -i\frac{\sqrt{2\pi}}{2}\text{sgn}(\omega). \tag{2.2}$$

Formulas of (2.2) are useful to derive the original Tretiak-Metz inversion formula. Assume that readers are knowledgeable of the principal-valued integral and the generalized distribution theory. Below I collect some well-known facts in the literature without detailed proofs. Most of these facts have been well established in many math and engineering textbooks.

**Special functions**. Three equivalent definitions of *n*-th order Bessel function of the first kind for integer *n* are presented below,

$$J_n(r) = \frac{1}{i^n 2\pi} \int_0^{2\pi} e^{i(n\theta + r\cos\theta)} d\theta = \frac{1}{\pi} \int_0^{\pi} \cos(n\theta - r\sin\theta) d\theta = \frac{1}{i^n \pi} \int_0^{\pi} e^{ir\cos\theta} \cos(n\theta) d\theta. \tag{2.3}$$

The Hankel transform and its inverse for integer *n* are expressed as:

$$H_n(\rho) = \int_0^{\infty} h(r) J_n(r\rho) r dr, \quad h(r) = \int_0^{\infty} H_n(\rho) J_n(r\rho) \rho d\rho. \tag{2.4}$$

We recall two important equalities from [13, 61],

$$\lim_{B \to \infty} \int_0^B J_0(|\vec{r}|\rho) \rho d\rho = \delta(\vec{r}), \tag{2.5}$$

$$\frac{1}{2\pi}\int_0^{2\pi} e^{a\sin\theta + i(n\theta + b\cos\theta)} d\theta = i^n \left(\frac{a+b}{\sqrt{b^2-a^2}}\right)^n J_n(\sqrt{b^2-a^2}) \text{ for } |b| > |a|. \quad (2.6)$$

**Singular integrals.** Let $\vec{\varphi} = (\cos\varphi, \sin\varphi)$ and $\vec{\varphi}^\perp = (-\sin\varphi, \cos\varphi)$. In the sense of principal integral value, from [31] we have

$$\frac{1}{2\pi}\int_0^{2\pi} \frac{\vec{\theta}}{\vec{\theta}\cdot\vec{\varphi}} e^{in\theta} d\theta = \begin{cases} 0, & n \text{ is odd} \\ \vec{\varphi}, & n = 0 \\ -i\,\text{sgn}(n) e^{in(\varphi-\pi/2)}\vec{\varphi}^\perp, & n \text{ is even.} \end{cases} \quad (2.7)$$

In the sense of generalized distribution, we have the following equality

$$\frac{1}{2\pi}\text{div}\frac{\vec{r}}{|\vec{r}|^2} = \delta(\vec{r}). \quad (2.8)$$

**Inversion of the cosh-weighted Hilbert transform.** More details on this topic can be found from [38-42]. The cosh-weighted Hilbert transform $H_\mu(s)$ of a function $h(t)$ in $I_q = (-q, q)$ is defined as

$$H_\mu(s) = \frac{1}{\pi}\int_{-q}^{q} \frac{\cosh(\mu(s-t))}{s-t} h(t)dt. \quad (2.9)$$

Define the weighted Hilbert space $L_w^2(I_q)$ as

$$L_w^2(I_q) = \{f(t) \in L^2(I_q) : \int_{-q}^{q} f^2(t)/\sqrt{q^2-t^2}\,dt < \infty\}. \quad (2.10)$$

It was proved in [42] that the cosh-weighted Hilbert transform is coercive in $L_w^2(I_q)$ so that there exists one and only one solution $h(t)$ for a given $H_\mu(s)$ satisfying the range condition. Furthermore, $h(t)$ can be resolved by the following iteration

$$h^{(n+1)}(t) = \int_{-q}^{q} \frac{H_\mu(s)}{\pi(s-t)\cosh(\mu s)}\sqrt{\frac{q^2-t^2}{q^2-s^2}}ds + \int_{-q}^{q} \frac{\tanh(\mu s)}{\pi(s-t)}\sqrt{\frac{q^2-t^2}{q^2-s^2}}\left[\int_{-q}^{q} \frac{\tanh(\mu u)h^{(n)}(u)}{\pi(s-u)}du\right]ds. \quad (2.11)$$

Based on the analysis in [42], sequence $h^{(n)}$ converges to $h^{(\infty)}$ in $L_w^2(I_q)$ and $h(t) = h^{(\infty)}(t)/\cosh(\mu t)$. However, the numerical realization of (2.11) may not be efficient in practical application, a more simple and stable matrix-inversion based method will be elaborated later.

**Two Hilbert spaces.** Let $L^2(D)$ be the Hilbert space of square-integrable functions in the unit disk $D$ with the norm,

$$||f(x,y)||_{L^2(D)}^2 = \iint_D f^2(x,y)dxdy. \quad (2.12)$$

For real number $\alpha \in R^1$, denote by $H_0^\alpha(S^1 \times (-1, 1))$ the Sobolev space of functions with compact support in $S^1 \times (-1, 1)$ under the following norm,

$$||g(\theta, s)||_{H_0^\alpha} = \left[\int_0^{2\pi} d\theta \int_{-\infty}^{\infty} |\tilde{g}(\theta, \omega)|^2 \sqrt[\alpha]{1+|\omega|^2}\,d\omega\right]^{1/2}, \quad (2.13)$$

where $\tilde{g}(\theta, \omega)$ is the 1D FT of $g(\theta, s)$ with respect to the second variable. It has been known that $\mathbf{R}_\mu$ is a bounded operator from $L^2(D)$ to $H_0^{0.5}(S^1 \times (-1, 1))$, and $\mathbf{R}_\mu^*$ is a bounded operator from $H_0^{-0.5}(S^1 \times (-1, 1))$ to $L^2(D)$. These functional results can be derived using the arguments in [62].

**Central slice theorem for the ERT.** The central slice theorem for the attenuation-free projection can be simply expressed, for real pairs $(\theta, \omega)$, as

$$\tilde{p}(\theta, \omega) = \tilde{f}(\omega\cos\theta, \omega\sin\theta). \quad (2.14)$$

With the presence of attenuation, relation (2.14) no longer holds. Let $C$ denote the complex plane, and $C^2$ is the product of two complex planes. From the profound Paley-Wiener theorem, the Fourier transform $\tilde{f}(\xi_x, \xi_y)$, when restricted on $R^2$, can be extended as an entire function $\tilde{f}(z_1, z_2)$ of $C^2$. Then for the ERT, we have the following generalized central slice theorem

$$\tilde{p}(\theta,\omega) = \tilde{f}(\omega\cos\theta + i\mu\sin\theta, \omega\sin\theta - i\mu\cos\theta). \tag{2.15}$$

Unlike the attenuation-free case in which the central slice theorem readily leads to the FBP algorithm through the relation between the polar coordinates and Cartesian coordinates, the FT $\tilde{f}(\xi_x,\xi_y)$ on $R^2$ is not immediately available from (2.15) because $(\omega\cos\theta + i\mu\sin\theta, \omega\sin\theta - i\mu\cos\theta)$ is not yet the Euclidean space $R^2$ when $(\theta,\omega)$ varies on $[0, 2\pi)\times R^1$. Thus, estimating $\tilde{f}(\xi_x,\xi_y)$ from $\tilde{p}(\theta,\omega)$ to reconstruct $f(x,y)$ by the inverse FT still remains unsolved.

For the case of $\mu \in R^1$, heuristically treating $C^2$ like $R^2$ in the works such as [21], one may use the polar coordinates of $C^2$ to calculate $\tilde{f}(\xi_x,\xi_y)$ from $\tilde{p}(\theta,\omega)$. It was surprising that the same inversion formula as the one in [13] could have been derived by such pseudo-math operation. However, I would like to clearly repeat that these ideas are fantastic possibly from the engineering point of view, but they are far fetched from a mathematical point of view because much more complicated several complex variables analysis might be needed to justify these pseudo-math operations. Here I cite comments from the recent review by Kuchment in [63]. For $(\theta,\omega) \in S^1 \times R^1$, let $S_\mu = \{\omega\theta - i\mu\theta^\perp\}$ be a smooth surface in $C^2$. As described in [63], function $\tilde{f}$ is known on $S_\mu$, the idea to use Cauchy type argument to shift the integration from $R^2$ to $S_\mu$ is not straightforward since, in particular, the surface $S_\mu$ has a hole in it and this is not holomogical to $R^2$. More details about the properties of several complex variables and the range condition of the ERT can be found in [64-69]. In recent works, *e.g.*, [36, 37], the same heuristic arguments of using the polar expression $\tilde{f}(\rho,\alpha)$ of $\tilde{f}(z_1,z_2)$ was used to handle the half-scan data reconstruction. It is unclear how these pseudo-math arguments can be justified using the several complex variables analysis.

For the case of $\mu = i\eta$ with $\eta \in R^1$, (2.15) becomes

$$\tilde{p}(\theta,\omega) = \tilde{f}(\omega\cos\theta - \eta\sin\theta, \omega\sin\theta + \eta\cos\theta). \tag{2.16}$$

Now (2.16) implies that $\tilde{f}(\xi_x,\xi_y)$ is only available for $|\xi_x|^2 + |\xi_y|^2 \geq \eta^2$. An explicit inversion of (2.16) is not obvious either. Coincidently, this issue has not been paid much attention in the literature except the work [8]. In this paper, we intend to extend most of the existing inversion formulas to complex-valued $\mu$.

In this section, we only list some necessary materials used in the paper. More details on the mathematical preliminary and application domain knowledge on the image reconstruction can be found from those popular textbooks for example [70-77].

## III. The Inverse Exponential Radon Transform

In this section, we shall survey the proofs of various inversion formulas derived in the literature. I try to make the proof as self-complete as possible without using mathematical results other than the facts stated in the preceding section, though some necessary details for non-expert readers may still need to be referred to the original papers. In particular, I give a very simple proof of the Tretiak-Metz-Novikov FBP-type inversion formula using the arguments by Natterer in [31]. The Fourier and CHE inversion formulas then can be easily derived from the FBP formula. On the other hand, readers should bear in mind that some inversion formulas may be only available for real-valued $\mu$ due to the limitation of derivation method.

### A. The FBP-type inversion formula

The most beautiful reconstruction algorithm unquestionably is the FBP procedure in [12] because of its theoretical simplicity and numerical efficiency. The first FBP reconstruction formula for the IERT was derived by Tretiak-Metz in [13]. Recently, the method by Novikov in [30] for the inverse attenuated Radon transform also leads to a similar FBP formula for the IERT, and easily implies the existence of an FBP-type algorithm for fan-beam data. Hereafter, $p'(\theta,l)$ stands for the derivative of $p(\theta,l)$ with respect to the second variable. We summarize these formulas into three flavors:

$$f(\vec{r}) = \frac{1}{4\pi^2} \int_0^{2\pi} e^{-\mu \vec{r}\cdot\vec{\theta}^\perp} [\frac{\partial}{\partial s} \int_{-\infty}^{\infty} \frac{e^{i\mu(s-l)}}{s-l} p(\theta,l)dl]|_{s=\vec{r}\cdot\vec{\theta}} d\theta$$

$$= \frac{1}{4\pi^2} \int_0^{2\pi} e^{-\mu \vec{r}\cdot\vec{\theta}^\perp} [\frac{\partial}{\partial s} \int_{-\infty}^{\infty} \frac{e^{-i\mu(s-l)}}{s-l} p(\theta,l)dl]|_{s=\vec{r}\cdot\vec{\theta}} d\theta \qquad (3.1)$$

$$= \frac{1}{4\pi^2} \int_0^{2\pi} e^{-\mu \vec{r}\cdot\vec{\theta}^\perp} [\frac{\partial}{\partial s} \int_{-\infty}^{\infty} \frac{\cos(\mu(s-l))}{s-l} p(\theta,l)dl]|_{s=\vec{r}\cdot\vec{\theta}} d\theta.$$

I want to mention that (3.1) takes the form in [35], which is slightly different from the expressions in other existing works [30-34]. Notice that the order of the partial derivative and convolution in (3.1) can be exchanged. Actually by (2.2), in the frequency domain, the convolution for $\mu > 0$ in (3.1) can be understood as

$$\frac{1}{\pi} \int_{-\infty}^{\infty} \frac{1}{s-l} \frac{\cos(\mu(s-l))}{s-l} p'(\theta,l)dl = \int_{|\omega|>\mu} |\omega| \tilde{p}(\theta,\omega) e^{i\omega s} d\omega. \qquad (3.2)$$

With (3.2) in mind, it is straightforward that the inversion formula (3.1) should be equivalent to the inversion formulas in [13, 30, 31] for real $\mu$. Using (2.7), we shall give a more elementary proof of (3.1). Notice that the terms with odd integers in (2.7) cancel out. Let $\vec{v} = (i, -1)$, we have

$$\int_0^{2\pi} \frac{\vec{\theta}}{\vec{\theta}\cdot(\vec{r}-\vec{w})} \exp(\mu\vec{v}\cdot(\vec{w}-\vec{r})e^{-i\theta})d\theta = \int_0^{2\pi} \frac{\vec{\theta}}{\vec{\theta}\cdot(\vec{r}-\vec{w})} \cosh(\mu\vec{v}\cdot(\vec{w}-\vec{r})e^{-i\theta})d\theta. \qquad (3.3)$$

Let $\vec{\varphi} = (\vec{r}-\vec{w})/|\vec{r}-\vec{w}| = (\cos\varphi, \sin\varphi)$ and $\vec{\varphi}^\perp = (-\sin\varphi, \cos\varphi)$, we derive the following relation

$$\frac{1}{2\pi} \int_0^{2\pi} \frac{\vec{\theta}}{\vec{\theta}\cdot(\vec{r}-\vec{w})} [\cosh(\mu\vec{v}\cdot(\vec{w}-\vec{r})e^{-i\theta}) - 1]d\theta = \frac{1}{i} \frac{\vec{\varphi}^\perp}{|\vec{r}-\vec{w}|} [\cosh(\mu\vec{v}\cdot(\vec{w}-\vec{r})e^{-i(\varphi-\pi/2)}) - 1]$$

$$= \frac{1}{i} \frac{\vec{\varphi}^\perp}{|\vec{r}-\vec{w}|} [\cosh(-\mu|\vec{r}-\vec{w}|) - 1]. \qquad (3.4)$$

We observe that $\text{div}\{\vec{\omega}^\perp[\cosh(a|\vec{r}-\vec{w}|)/|\vec{r}-\vec{w}|]\} = [\vec{\omega}^\perp \cdot \nabla][\cosh(a|\vec{r}-\vec{w}|)/|\vec{r}-\vec{w}|] = 0$, then by (3.3) and (3.4) we have

$$\text{div} \int_0^{2\pi} \frac{\vec{\theta}}{\vec{\theta}\cdot(\vec{r}-\vec{w})} e^{(-\mu\vec{\theta}^\perp + i\mu\vec{\theta})\cdot(\vec{w}-\vec{r})} d\theta = \text{div} \int_0^{2\pi} \frac{\vec{\theta}}{\vec{\theta}\cdot(\vec{r}-\vec{w})} \exp(\mu\vec{v}\cdot(\vec{w}-\vec{r})e^{-i\theta})d\theta$$

$$= \text{div} \int_0^{2\pi} \frac{\vec{\theta}}{\vec{\theta}\cdot(\vec{r}-\vec{w})} \cosh(\mu\vec{v}\cdot(\vec{w}-\vec{r})e^{-i\theta})d\theta \qquad (3.5)$$

$$= \text{div} \int_0^{2\pi} \frac{\vec{\theta}}{\vec{\theta}\cdot(\vec{r}-\vec{w})} d\theta$$

$$= 2\pi \, \text{div} \frac{(\vec{r}-\vec{w})}{|\vec{r}-\vec{w}|^2}.$$

The derivation of (3.5) is based on the arguments from [31, 35]. Notice that $[\vec{\theta}\cdot\nabla]e^{-\mu\vec{\theta}^\perp \cdot(\vec{r}-\vec{w})} = 0$, by using (2.8), one can express function $f(\vec{r})$ as

$$f(\vec{r}) = \frac{1}{4\pi^2} \iint [\text{div} \int_0^{2\pi} \vec{\theta} e^{-\mu\vec{\theta}^\perp \cdot(\vec{r}-\vec{w})} \frac{e^{i\mu\vec{\theta}\cdot(\vec{r}-\vec{w})}}{\vec{\theta}\cdot(\vec{r}-\vec{w})} d\theta] f(\vec{w}) d\vec{w}$$

$$= \frac{1}{4\pi^2} \int_0^{2\pi} [\iint e^{-\mu\vec{\theta}^\perp \cdot(\vec{r}-\vec{w})} [\vec{\theta}\cdot\nabla] \frac{e^{i\mu\vec{\theta}\cdot(\vec{r}-\vec{w})}}{\vec{\theta}\cdot(\vec{r}-\vec{w})} f(\vec{w}) d\vec{w}] d\theta. \qquad (3.6)$$

Let $l = \vec{\theta}\cdot\vec{w}$ and $t = \vec{\theta}^\perp \cdot \vec{w}$, then it is easy to verify the following expression

$$\frac{1}{4\pi^2}\int_0^{2\pi}e^{-\mu\vec{r}\cdot\vec{\theta}^\perp}[\int_{-\infty}^{\infty}\frac{\partial}{\partial s}\frac{e^{i\mu(s-l)}}{s-l}p(\theta,l)dl]|_{s=\vec{r}\cdot\vec{\theta}}\,d\theta$$

$$=\frac{1}{4\pi^2}\int_0^{2\pi}e^{-\mu\vec{r}\cdot\vec{\theta}^\perp}\{\int_{-\infty}^{\infty}[\vec{\theta}\cdot\nabla]\frac{e^{i\mu\vec{\theta}\cdot(\vec{r}-\vec{w})}}{\vec{\theta}\cdot(\vec{r}-\vec{w})}[\int_{-\infty}^{\infty}f(l\vec{\theta}+t\vec{\theta}^\perp)e^{\mu t}dt]dl\}d\theta \quad (3.7)$$

$$=\frac{1}{4\pi^2}\int_0^{2\pi}\{\iint e^{-\mu\vec{\theta}^\perp\cdot(\vec{r}-\vec{w})}[\vec{\theta}\cdot\nabla]\frac{e^{i\mu\vec{\theta}\cdot(\vec{r}-\vec{w})}}{\vec{\theta}\cdot(\vec{r}-\vec{w})}f(\vec{w})d\vec{w}\}d\theta$$

$$=f(\vec{r}).$$

Therefore inversion formula (3.1) has been proven through combining equations from (3.5) to (3.7). The second line of (3.1) follows the same reason. Then it follows that the third line of (3.1) holds. We also remark that the original proof of Theorem 1 in [13] used equation (2.5), and does not appear to be valid to the case of complex-valued $\mu$.

### B. The Fourier inversion method

The Fourier inversion method involves both the FT and FSE. Write the circular harmonic expansions of $\tilde{f}(\xi_x,\xi_y)$ and $f(x,y)$ as

$$\tilde{f}(\rho,\psi)=\sum_{-\infty}^{\infty}\tilde{f}_n(\rho)e^{in\psi} \text{ and } f(r,\varphi)=\sum_{-\infty}^{\infty}f_n(r)e^{in\varphi} \quad (3.8)$$

where $\rho=\sqrt{\xi_x^2+\xi_y^2}$, $\psi=\mathrm{atan}(\xi_x/\xi_y)$ and $\varphi=\mathrm{atan}(x/y)$. By the FT, coefficients $\tilde{f}_n(\cdot)$ and $f_n(\cdot)$ satisfy

$$f_n(r)=i^n\int_0^{\infty}\tilde{f}_n(\rho)J_n(r\rho)\rho d\rho \text{ and } \tilde{f}_n(\rho)=(-i)^n\int_0^{\infty}f_n(r)J_n(r\rho)rdr. \quad (3.9)$$

Similarly, write the FSE of $\tilde{p}(\theta,\omega)$ with respect to variable $\theta$ as

$$\tilde{p}(\theta,\omega)=\sum_{-\infty}^{\infty}\tilde{p}_n(\omega)e^{in\theta}. \quad (3.10)$$

For the case of $\omega,\mu\in R^1$ with $|\omega|>|\mu|$, the following relation between $\tilde{f}_n(\cdot)$ and $\tilde{p}_n(\omega)$ was originally derived in [13, 17],

$$\tilde{f}_n(\sqrt{\omega^2-\mu^2})=\frac{(\omega^2-\mu^2)^{n/2}}{(\omega-\mu)^n}\tilde{p}_n(\omega)=(-1)^n\frac{(\omega^2-\mu^2)^{n/2}}{(\omega+\mu)^n}\tilde{p}_n(-\omega). \quad (3.11)$$

Notice that (3.11) implies the following symmetry property,

$$(\mu+\omega)^n\tilde{p}_n(\omega)=(\mu-\omega)^n\tilde{p}_n(-\omega) \text{ if } |\omega|>|\mu|. \quad (3.12)$$

As revealed in [13], the derivation of (3.11) from (2.6) and (3.9) is rather straightforward. Without loss of generality, we assume $\mu>0$. By variable change $l=r\cos\varphi$ and $t=r\sin\varphi$, we derive

$$\int_0^{2\pi}e^{-in\theta}d\theta\int_{-\infty}^{\infty}[\int_{-\infty}^{\infty}f(l\vec{\theta}+t\vec{\theta}^\perp)e^{\mu t}dt]e^{-i\omega l}dl$$

$$=\int_0^{2\pi}e^{-in\theta}d\theta\int_0^{\infty}\int_0^{2\pi}f(r\cos(\theta+\varphi),r\sin(\theta+\varphi))e^{-i\omega r\cos\varphi+\mu r\sin\varphi}rdrd\varphi$$

$$=\int_0^{\infty}\{\int_0^{2\pi}[\int_0^{2\pi}e^{-in(\theta+\varphi)}f(r\cos(\theta+\varphi),r\sin(\theta+\varphi))d\theta]e^{in\varphi-i\omega r\cos\varphi+\mu r\sin\varphi}d\varphi\}rdr \quad (3.13)$$

$$=\int_0^{\infty}[\int_0^{2\pi}e^{in\varphi+i\omega r\cos\varphi+\mu r\sin\varphi}d\varphi]f_n(r)rdr.$$

From (2.6) and (3.9), we have

$$\begin{aligned}
\tilde{p}_n(\omega) &= \frac{1}{2\pi} \int_0^{2\pi} e^{-in\theta} d\theta \int_{-\infty}^{\infty} p(\theta,l) e^{-i\omega l} dl \\
&= \frac{1}{2\pi} \int_0^{2\pi} e^{-in\theta} d\theta \int_{-\infty}^{\infty} [\int_{-\infty}^{\infty} f(l\vec{\theta}+t\vec{\theta}^{\perp}) e^{\mu t} dt] e^{-i\omega l} dl \\
&= i^n (\frac{\mu-\omega}{\sqrt{\omega^2-\mu^2}})^n \int_0^{\infty} J_n(r\sqrt{\omega^2-\mu^2}) f_n(r) r dr \\
&= (\frac{\omega-\mu}{\sqrt{\omega^2-\mu^2}})^n \tilde{f}_n(\sqrt{\omega^2-\mu^2}).
\end{aligned} \quad (3.14)$$

Take $\pm\omega$ in (3.14), we derive (3.11). The derivations of (3.1) and (3.11) are independent. In particular, formula (3.12) reveals certain redundancy in the projection data. This may be useful to the design of optimal filters to reduce the noise in the reconstructed image as investigated in [22]. Remember that there are three different expressions in (3.1). Actually, the first two expressions correspond to (3.14) for the positive and negative $\omega$. Next we derive the associations between (3.1) and (3.14). Recall the FT of $e^{i\mu l}$ is $\sqrt{2\pi}\delta(\omega-\mu)$. In the polar coordinates, the first line of (3.1) becomes

$$\begin{aligned}
f(r,\varphi) &= \frac{1}{4\pi} \int_0^{2\pi} e^{\mu r \sin(\theta-\varphi)} \{\int_{\omega>\mu} \omega \tilde{p}(\theta,\omega) e^{i\omega r \cos(\theta-\varphi)} d\omega\} d\theta \\
&= \frac{1}{4\pi} \int_0^{2\pi} e^{\mu r \sin(\theta-\varphi)} \{\int_{\omega>\mu} \omega [\sum_{-\infty}^{\infty} \tilde{p}_n(\omega) e^{in\theta}] e^{i\omega r \cos(\theta-\varphi)} d\omega\} d\theta \\
&= \frac{1}{4\pi} \sum_{-\infty}^{\infty} e^{in\varphi} \int_{\omega>\mu} \omega \tilde{p}_n(\omega) [\int_0^{2\pi} e^{\mu r \sin\theta + in\theta + i\omega r \cos\theta} d\theta] d\omega \\
&= \sum_{-\infty}^{\infty} e^{in\varphi} [\frac{1}{2} \int_{\omega>\mu} i^n (\frac{\omega-\mu}{\sqrt{\omega^2-\mu^2}})^n J_n(r\sqrt{\omega^2-\mu^2}) \omega \tilde{p}_n(\omega) d\omega].
\end{aligned} \quad (3.15)$$

With variable change $\omega = \sqrt{\rho^2+\mu^2}$, rewrite the inner integral in the last line of (3.13) as

$$\int_{\mu}^{\infty} i^n (\frac{\omega-\mu}{\sqrt{\omega^2-\mu^2}})^n \tilde{p}_n(\omega) J_n(r\sqrt{\omega^2-\mu^2}) \omega d\omega = \int_0^{\infty} i^n (\frac{\sqrt{\rho^2+\mu^2}-\mu}{\rho})^n \tilde{p}_n(\sqrt{\rho^2+\mu^2}) J_n(r\rho) \rho d\rho. \quad (3.16)$$

Comparing (3.16) and (3.9), we derive the first equality in (3.11). Thus expressions in (3.1) represent three different weighting methods of (3.11). In this sense, formulas (3.1) and (3.11) can be mutually deduced. More detailed analysis of the various weighting methods can be found from [22].

For complex $\mu$, the preceding arguments using the FSE do not appear to hold. Especially, for $\mu = i\eta$ with $\eta \in R^1$, equality (3.13) does not imply any method to reconstruct $f(\vec{r})$ from $p(\theta,l)$. Anyway, we include that relation below,

$$\tilde{p}_n(\omega) = \tilde{f}_n(\sqrt{\omega^2+\eta^2}) e^{-in\psi_\omega}, \quad (3.17)$$

here $\psi_\omega = \arcsin(\eta/\sqrt{\omega^2+\eta^2})$.

**Remark 1**. I would like to comment on the use of relations (3.11, 3.12). In history, some works, for example [20-22], suggested other ways of using (3.11) by weighting two redundant terms differently. In the numerical implementation, the nonlinear transformation $\rho = \sqrt{\omega^2-\mu^2}$ requires the interpolation in the Fourier domain. This could introduce the extra blurring in the reconstructed image.

### C. The circular harmonic expansion method

Write the circular harmonic expansion of $f(x,y)$ as

$$f(r,\varphi) = \sum_{-\infty}^{\infty} f_n(r) e^{in\varphi}. \quad (3.18)$$

Similarly, write the Fourier series expansion of $p(\theta,s)$ as

$$p(\theta,s) = \sum_{-\infty}^{\infty} p_n(s)e^{in\theta}. \tag{3.19}$$

By variable change $t = \pm\sqrt{r^2 - s^2}$ and $\varphi = \mathrm{acos}(s/r)$, we express $p_n(s)$ for $s > 0$ in an integral transform of $f_n(r)$ as

$$\begin{aligned}
p_n(s) &= \int_0^{2\pi}\int_{-\infty}^{\infty} f(s\vec{\theta} + t\vec{\theta}^\perp)e^{\mu t}dt\, e^{-in\theta}d\theta \\
&= 2\int_s^{\infty}\int_0^{2\pi}[f(r,\theta)e^{-in\theta}d\theta]e^{\mu\sqrt{r^2-s^2}}e^{-in[\mathrm{acos}(s/r)]}\frac{r}{\sqrt{r^2-s^2}}dr \\
&= 2\int_s^{\infty} e^{\mu\sqrt{r^2-s^2}} e^{-in[\mathrm{acos}(s/r)]}\frac{f_n(r)r}{\sqrt{r^2-s^2}}dr
\end{aligned} \tag{3.20}$$

When $\mu \equiv 0$, a closed-form formula to calculate $p_n(s)$ from $f_n(r)$ was derived in [11] through using the Chebyshev polynomials. That formula was extended to the imaginary exponential constant $\mu = i\eta$ in [26]. It was mentioned in [26] that Cormack-type inversion formula should be available for real $\mu$. Here we give a proof for a complex constant. Rewrite the second line of (3.1) as follows

$$\begin{aligned}
f(r,\varphi) &= \frac{1}{4\pi^2}\int_0^{2\pi} e^{-\mu\vec{r}\cdot\vec{\theta}^\perp}[\int_{-\infty}^{\infty}\frac{e^{-i\mu(s-l)}}{s-l}p'(\theta,l)dl]|_{s=\vec{r}\cdot\vec{\theta}}\,d\theta \\
&= \frac{1}{4\pi^2}\int_0^{2\pi} e^{\mu r\sin(\theta-\varphi)}[\int_{-\infty}^{\infty}\frac{e^{-i\mu(r\cos(\theta-\varphi)-l)}}{r\cos(\theta-\varphi)-l}p'(\theta,l)dl]d\theta \\
&= \frac{1}{4\pi^2}\int_0^{2\pi} e^{\mu r\sin(\theta-\varphi)}[\int_{-\infty}^{\infty}\frac{e^{-i\mu(r\cos(\theta-\varphi)-l)}}{r\cos(\theta-\varphi)-l}\sum_{-\infty}^{\infty}p'_n(l)e^{in\theta}dl]d\theta \\
&= \frac{1}{4\pi^2}\sum_{-\infty}^{\infty} e^{in\varphi}\int_{-\infty}^{\infty}[\int_0^{2\pi}\frac{e^{\mu r\sin\theta - i\mu(r\cos\theta-l)}}{r\cos\theta-l}e^{in\theta}d\theta]p'_n(l)dl \\
&= \frac{1}{4\pi^2}\sum_{-\infty}^{\infty} e^{in\varphi}\int_{-\infty}^{\infty}[\int_0^{2\pi}\frac{\exp(-i\mu re^{i\theta}+l+in\theta)}{r\cos\theta-l}d\theta]p'_n(l)dl.
\end{aligned} \tag{3.21}$$

Let $K_n(r,l)$ denote the inner integral in the last line of (3.22), according to the evaluation of [26, eqn (4.4)], we have a closed-form expression of $K_n(r,l)$ for $|l| > r$ as follows

$$K_n(r,l) = -\frac{2\pi}{\sqrt{l^2-r^2}}\exp[i\mu\sqrt{l^2-r^2}\,][\frac{|l|-\sqrt{l^2-r^2}}{r}\mathrm{sgn}(l)]^n. \tag{3.23}$$

Then we have the explicit formula

$$f_n(r) = -\frac{1}{2\pi}\int_{|l|>r}\exp[i\mu\sqrt{l^2-r^2}\,][\frac{|l|-\sqrt{l^2-r^2}}{r}\mathrm{sgn}(l)]^n\frac{p'_n(l)}{\sqrt{l^2-r^2}}dl. \tag{3.23}$$

Notice that (3.23) is not a one-step closed-form expression yet. Some simplified closed-form expression of the inner integral in the right-hand side of (3.23) can be found in [26]. As suggested in [25], the fast Fourier transform can be used to calculate the Fourier series coefficients of $e^{\mu r\sin\theta + i\mu(r\cos\theta-l)}/(r\cos\theta - l)$ in a numerically more stable and efficient way, thus, a closed-form expression of $f_n(r)$ may not be necessary in the numerical realization. I want to comment on the Fourier inversion and circular harmonic expansion methods. Because of the FT and FSE, the sampling rate of the projection data has to be the power of 2 in order to use the FFT. However, the clinic SPECT data rarely meets such condition. So far, the only advantage of (3.23) over (3.1) seems to be the easy-to-use in handling the variable-focal-length fan-beam projection data to avoid rebinning projection data [25].

### D. The DBH method

The DBH reconstruction method was developed by Rullgård in [38] for the IERT with 180º data. That inversion method seems to be quite different from the preceding FBP inversion formula and other equivalents. We shall explore more properties of the DBH method. First we define an intermediate function $\hat{f}(x,y)$ as follows:

$$\hat{f}(x,y) = -\frac{1}{2\pi} \int_{-\pi/2}^{\pi/2} e^{-\mu \vec{r} \cdot \vec{\theta}^{\perp}} p'(\theta,s)|_{s=\vec{r}\cdot\vec{\theta}} \, d\theta \, . \tag{3.22}$$

The right-hand side of (3.22) is commonly called the DBP (derivative and back-projection) operation since it is composed of two steps: one is the partial derivative of the attenuated projection data $p(\theta,s)$ and the other is the weighted back-projection. Express $p'(\theta,s)$ as:

$$p'(\theta,s) = \int_{-\infty}^{\infty} [\vec{\theta} \cdot \nabla f)](s\vec{\theta}+t\vec{\theta}^{\perp})e^{\mu t} dt \, . \tag{3.23}$$

With the change of variables $\tau = t - \vec{r} \cdot \vec{\theta}^{\perp}$ and the order exchange of iterated integrals, one derives a fundamental relation between $\hat{f}(x,y)$ and $f(x,y)$ as follows:

$$\begin{aligned}
\hat{f}(x,y) &= -\frac{1}{2\pi} \int_{-\pi/2}^{\pi/2} \{ \int_{-\infty}^{\infty} [\vec{\theta} \cdot \nabla f)](s\vec{\theta}+t\vec{\theta}^{\perp})e^{\mu(t-\vec{r}\cdot\vec{\theta}^{\perp})} dt \}|_{s=\vec{r}\cdot\vec{\theta}} \, d\theta \\
&= -\frac{1}{2\pi} \int_{-\infty}^{\infty} e^{\mu\tau} d\tau \int_{-\pi/2}^{\pi/2} [\vec{\theta} \cdot \nabla f](x-\tau\sin\theta, y+\tau\cos\theta) d\theta \\
&= \frac{1}{2\pi} \int_{-\infty}^{\infty} \frac{e^{\mu\tau} d\tau}{\tau} \int_{-\pi/2}^{\pi/2} [\frac{\partial}{\partial \theta} f](x-\tau\sin\theta, y+\tau\cos\theta) d\theta \\
&= \frac{1}{2\pi} \int_{-\infty}^{\infty} \frac{e^{\mu\tau}[f(x-\tau,y)-f(x+\tau,y)]}{\tau} d\tau \\
&= \int_{-\infty}^{\infty} \frac{\cosh(\mu\tau)}{\pi\tau} f(x-\tau,y) d\tau = \int_{-\infty}^{\infty} \frac{\cosh(\mu(x-\tau))}{\pi(x-\tau)} f(\tau,y) d\tau.
\end{aligned} \tag{3.24}$$

Relation $[\partial f / \partial \theta](x-\tau\sin\theta, y+\tau\cos\theta) = -\tau[\vec{\theta} \cdot \nabla f](x-\tau\sin\theta, y+\tau\cos\theta)$ is used in the derivation of (3.24). The last line of (3.24) indicates that the result of the DBP operation is the cosh-weighted Hilbert transform of $f(x,y)$ along the horizontal lines. This implies that the reconstruction of $f(x,y)$ from $p(\theta,s)$ can be carried out through a two-step processing: one is the DBP operation, and the other is the inversion of the cosh-weighted Hilbert transform.

**Remark 2**. In the calculations from (3.22) to (3.24), if we choose attenuation constant $\mu = i\eta$ as a pure imaginary number, the resulting weight in (3.24) becomes $\cos(\eta(\tau-x))$. In the frequency domain, the effect of kernel $\cos(\eta(\tau-x))$ is equivalent to removing the low frequency component of $|\omega| \leq \eta|$. The imaginary attenuation does have several applications as described in [7-9].

Notice that integral interval in (3.24) is $(-\infty, \infty)$. If the support of $f(x,y)$ on that line is $[-L(y), L(y)]$, then the inversion formula (2.11) can be applicable to reconstruct $f(x,y)$ from $\hat{f}(x,y)$. The detailed numerical realization of the reconstruction procedure will be discussed later. The assumption that $f(x,y)$ has support in interval $[-L(y), L(y)]$ is very crucial in order to use (2.11). An illustration of the back-projection area and the line segment for the finite inversion of the weighted Hilbert transform is shown in Fig 2 for application of the SPECT imaging. Two advantages in the DBP operation include the less data required in the backprojection over 180º range and the adaptability to handle other more complicated data acquisition geometries such as fan-beam or variable-focal-length fan-beam. So far we have assumed that the backprojection is performed over $[-\pi/2, \pi/2]$. For $\psi \in [0, \pi)$, define $\hat{f}_\psi(x,y)$ as

$$\hat{f}_\psi(x,y) = -\frac{1}{2} \int_{\psi-\pi/2}^{\psi+\pi/2} e^{-\mu\vec{r}\cdot\vec{\theta}^{\perp}} p'(\theta, \vec{r} \cdot \vec{\theta}) d\theta \, . \tag{3.25}$$

One can easily derive the following relations:

$$\hat{f}_\psi(r\cos\psi, r\sin\psi) = \int_{-\infty}^{\infty} \frac{\cosh(\mu(r-\tau)) f(\tau\cos\psi, \tau\sin\psi)}{r-\tau} d\tau \, , \tag{3.26}$$

$$\hat{f}_\psi(r\cos\psi, r\sin\psi) = -\hat{f}_{\psi+\pi}(-r\cos\psi, -r\sin\psi). \tag{3.27}$$

Then reconstruction of $f(x,y)$ from $p(\theta,s)$ reduces to the inversion of $\hat{f}_\psi(r\cos\psi, r\sin\psi)$ to obtain $f(\tau\cos\psi, \tau\sin\psi)$ for $\psi \in [0,\pi)$ and $\tau \in (-1,1)$ in the polar coordinates. If $p(\theta,s)$ is available over $[0, 2\pi]$, the asymmetry relation (3.27) can be used to mitigate the reconstruction noise since averaging two random variables with the same variance could reduce the resulting variance.

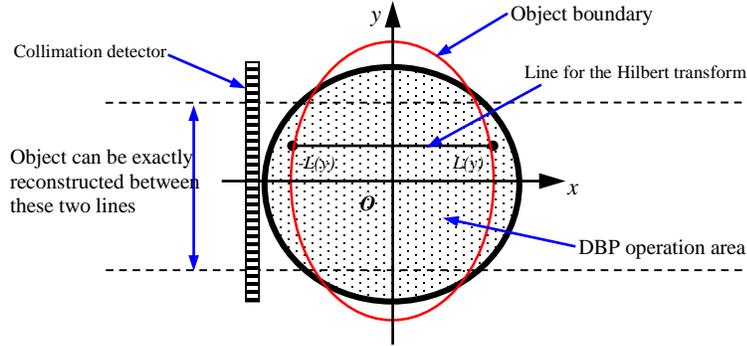

Fig 2: The radionuclide tracer is distributed inside the red ellipse. The collimation detector may not cover the entire region that contains the tracer. The DBP operation can be performed in the FOV of the detector inside the black thick circle. [–L(y), L(y)] is the finite interval contained in the overlapping area of FOV and the object boundary. The inversion of the weighted Hilbert transform is carried out on that interval for each y.

### E. Other analytical methods

One interesting flavor of (3.1) was mentioned in [13] as follows,

$$\begin{aligned}
f(0) &= -\frac{1}{4\pi^2} \int_0^{2\pi}\int_{-\infty}^{\infty} \frac{\cos(\mu l)}{l} p'(\theta,l) dl d\theta \\
&= \int_0^{\infty} \frac{\cos(\mu l)}{\pi l} \{\frac{1}{4\pi}\int_0^{2\pi} [p'(\theta,-l) - p'(\theta,l)]d\theta\} dl.
\end{aligned} \tag{3.28}$$

Formula (3.28) for the case of $\mu = 0$ was actually first derived by Radon in his historic paper [1], and was Corollary 2 of [13]. Recently, Puro used (3.28) to derive an explicit CHE formula in [26].

Another different procedure of back-projection followed by a 2D filtering was considered for the **IERT** in the early work [27]. Recently, further investigations about that idea were reported in [28, 29]. We calculate the back-projection operator $\mathbf{R}^*_\mu$ directly on $p(\theta,s)$ to obtain $\bar{f}(x,y)$ defined by

$$\begin{aligned}
\bar{f}(x,y) &= \frac{1}{2\pi}\int_0^{2\pi} e^{-\mu\vec{r}\cdot\vec{\theta}^\perp} p(\theta,s)|_{s=\vec{r}\cdot\vec{\theta}} d\theta \\
&= \frac{1}{2\pi}\int_0^{2\pi} \{\int_{-\infty}^{\infty} f(s\vec{\theta}+t\vec{\theta}^\perp)e^{\mu(t-\vec{r}\cdot\vec{\theta}^\perp)} dt\}|_{s=\vec{r}\cdot\vec{\theta}} d\theta \\
&= \frac{1}{2\pi}\int_{-\infty}^{\infty} e^{\mu\tau} d\tau \int_0^{2\pi} f(x-\tau\sin\theta, y+\tau\cos\theta) d\theta \\
&= \frac{1}{\pi}\int_0^{2\pi}\int_0^{\infty} \frac{\cosh(\mu\tau)}{\tau} f(x-\tau\sin\theta, y+\tau\cos\theta)\tau d\tau d\theta \\
&= \iint_{R^2} \frac{\cosh(\mu\sqrt{u^2+v^2})}{\pi\sqrt{u^2+v^2}} f(x-u, y-v) du dv \\
&= \iint_D \frac{\cosh(\mu\sqrt{(x-u)^2+(y-v)^2})}{\pi\sqrt{(x-u)^2+(y-v)^2}} f(u,v) du dv.
\end{aligned} \tag{3.29}$$

Then $f(x, y)$ can be reconstructed from $\bar{f}(x, y)$ through the deconvolution process. But no explicit inversion formula has ever been obtained. Also the back-projection in (3.29) needs to be performed in a larger area than the support of $f(x, y)$ in order to avoid too much loss of the high frequency component of $f(x, y)$. In history, this method never appeared to have a success although it looks simple either from the theoretical perspective or from numerical interests.

**Remark 3**. Equation (3.29) can be rewritten as the following operator in $L^2(D)$,

$$\bar{f}(\vec{r}) = \iint_D \frac{\cosh(\mu|\vec{r}-\vec{w}|)}{\pi|\vec{r}-\vec{w}|} f(\vec{w}) d\vec{w}. \tag{3.30}$$

In order to avoid the infinite support of the kernel function in (3.30), one may consider the finite inversion of (3.30) when restricted inside the disk $D$. Let $\chi_D$ be the characteristic function of the disk $D$, rewrite (3.30) as

$$\bar{f}(\vec{r}) = \chi_D(\vec{r}) \iint_{R^2} \frac{\cosh(\mu|\vec{r}-\vec{w}|)}{\pi|\vec{r}-\vec{w}|} \chi_D(\vec{w}) f(\vec{w}) d\vec{w}. \tag{3.31}$$

It's unclear whether a simple inversion for (3.31) exists. In this author's view, this may be an interesting problem from both mathematical interests and practical applications.

### F. Identification of real constant μ and range condition

Without specifically stating, we assume that $\mu$ is a real constant. If $p(\theta, s)$ is the ERT of certain function $f(x, y)$, then symmetry (3.12) holds. Actually, that condition is also sufficient for a function $g(\theta, s)$ of $S^1 \times (-1, 1)$ to be the ERT of certain function of $R^2$ as shown in [66, 67]. Furthermore, equation (3.12) can be used for the identification problem of estimating the real constant $\mu$ from projection data $p(\theta, s)$. Assume that $\tilde{p}_n(\omega) \neq \tilde{p}_n(-\omega)$ holds for at least one pair $(n, \omega)$ with $|\omega| > \mu$, then one can easily derive the following explicit formula

$$\mu = 2\omega\left[\frac{\sqrt[n]{\tilde{p}_n(-\omega)}}{\sqrt[n]{\tilde{p}_n(-\omega)} - \sqrt[n]{\tilde{p}_n(\omega)}} - \frac{1}{2}\right]. \tag{3.32}$$

This implies that the ERT is sufficient to determine the exponential constant $\mu$ and the original function $f(x, y)$ simultaneously as long as $f(x, y)$ is a non-radial function. More details on the identification problem can be found in [68, 69]. When $\mu = i\eta$ with $\eta \in R^1$, we observe $\tilde{p}_n(\omega) = \tilde{p}_n(-\omega)$, thus it does not seem to be possible to estimate the imaginary constant from the projection data $p(\theta, s)$ by (3.12).

For real-valued $\mu$, equation (3.12) is a necessary and sufficient condition for a function of $S^1 \times (-1, 1)$ to be the ERT of certain function of $R^2$. In the case of complex $\mu$, a different range condition was obtained in [35]. That new range condition is very similar to the one derived by Novikov in [30] for the attenuated Radon transform, which is expressed as an integral equation

$$\frac{1}{4\pi^2} \int_0^{2\pi} e^{-\mu\vec{r}\cdot\vec{\theta}^\perp} \int_{-\infty}^{\infty} \frac{\cos(\mu(\vec{r}\cdot\vec{\theta}-l))}{\vec{r}\cdot\vec{\theta}-l} p(\theta, l) dl d\theta \equiv 0. \tag{3.33}$$

The derivation of (3.33) follows the same arguments in the proof of (3.1) through the following equation

$$\frac{1}{2\pi}\int_0^{2\pi} \frac{1}{\vec{\varphi}\cdot\vec{\theta}} e^{in\theta} d\theta = \begin{cases} 0 & n \text{ is even} \\ i\,\text{sgn}(n) e^{in(\varphi-\pi/2)} & n \text{ is odd.} \end{cases} \tag{3.34}$$

We point out that (3.33) holds for complex constant $\mu$ and actually is equivalent to (3.12) if $\mu \in R^1$. Using the FT and FSE, we rewrite (3.33) as

$$\int_0^{2\pi} e^{-in\varphi} d\varphi \int_0^{2\pi} e^{\mu r \sin(\theta-\varphi)} \int_{|\omega|>\mu} \text{sgn}(\omega) \tilde{p}(\theta, \omega) e^{i\omega r \cos(\theta-\varphi)} d\omega d\theta \equiv 0. \tag{3.35}$$

Equation (3.35) is also equivalent to

$$\int\limits_{|\omega|>\mu} [\int\limits_0^{2\pi} e^{in\varphi+i\omega r\cos\varphi+\mu r\sin\varphi}d\varphi]\operatorname{sgn}(\omega)\tilde{p}_n(\omega)d\omega \equiv 0. \quad (3.36)$$

From (3.36), we derive

$$\int\limits_\mu^\infty [(\frac{\mu+\omega}{\sqrt{\omega^2-\mu^2}})^n \tilde{p}_n(\omega) - (\frac{\mu-\omega}{\sqrt{\omega^2-\mu^2}})^n \tilde{p}_n(-\omega)] J_n(r\sqrt{\omega^2-\mu^2})d\omega \equiv 0. \quad (3.37)$$

Changing variable $\omega = \sqrt{\rho^2+\mu^2}$, equality (3.37) becomes

$$\int\limits_0^\infty [(\frac{\mu+\sqrt{\rho^2+\mu^2}}{\rho})^n \tilde{p}_n(\sqrt{\rho^2+\mu^2}) - (\frac{\mu-\sqrt{\rho^2+\mu^2}}{\rho})^n \tilde{p}_n(-\sqrt{\rho^2+\mu^2})] \frac{1}{\sqrt{\rho^2+\mu^2}} J_n(r\rho)\rho d\rho \equiv 0. \quad (3.38)$$

By the inverse Hankel transform, we derive (3.12) from (3.38).

**Remark 4**. Equation (3.32) is explicit but only available to real-valued $\mu$. In the general case of complex $\mu$, identity (3.33) may be useful to estimate the attenuator $\mu$ as well through solving a linear equation system based on the discrete version of (3.33).

## IV. The IERT with Partial Measurements

In some applications such as cardiac SPECT, the projection data $p(\theta,s)$ is available over a subset of $[0, 2\pi]$. If the subset is $[0, \pi]$, the DBH method should be valid. Otherwise, these aforementioned algorithms do not provide an exact reconstruction for the partially available data. The partial data typically represents three different data formations: half-scan for $[0, \pi] \times (-1, 1)$, limited-angle for $[0, \rho] \times (-1, 1)$, here $0 < \rho < \pi$, and truncated-bin for $[0, 2\pi] \times (-\lambda, \lambda)$, here $0 < \lambda < 1$. Recent book [77] includes a summary about the exact reconstruction from those partial data, for instance, limited-angle and truncated-bin data if the attenuation is not present. Lately, an increasing interest has been brought up in those works [36-47] on the exact reconstruction from attenuated partial projection data. In these works, it confirms that the exact reconstruction from half-scan and certain truncated-bin data is possible, but an exact reconstruction from limited-angle data still remains unknown. In this paper, we will review these exact reconstruction methods for the partial data.

### A. Half-scan projection data

The integral in (3.24) with respect to $\theta$ is over $[-\pi/2, \pi/2]$. More general scanning range with total $\pi$ was considered in [37, 43, 47]. Denote by $\Lambda$ a subset of $[0, 2\pi]$ and $\Lambda^c$ its compliment. $\Lambda$ and $\Lambda^c$ satisfy

$$\Lambda = \bigcup_{1\leq i\leq N} \Lambda_i \text{ with } \Lambda_i \bigcap_{i\neq j} \Lambda_j = \emptyset, |\Lambda| = \pi, \Lambda = \pi + \Lambda^c \text{ and } N \text{ being a finite integer.} \quad (4.1)$$

Conditions of (4.1) indicate that $\Lambda$ is the union of finite number of non-overlapping intervals such that the total range is $\pi$ and two opposite rays are not in $\Lambda$ simultaneously. In this paper, the half-scan stands for the scanning range of $\theta$ over the subset $\Lambda$ defined in (4.1). First we give a unified analysis for several iterative methods in the literature. For any measurable subset $\Omega$ of $[0, 2\pi]$, we define $f_\Omega(\vec{r})$ as

$$f_\Omega(\vec{r}) = \frac{1}{4\pi^2} \int_\Omega e^{-\mu\vec{r}\cdot\vec{\theta}^\perp} [\int\limits_{-\infty}^\infty \frac{\exp(i\mu(s-l))}{s-l} p'(\theta,l)dl]|_{s=\vec{r}\cdot\vec{\theta}} \, d\theta. \quad (4.2)$$

Recall the coordinate transformation of $l = \vec{\theta}\cdot\vec{w}$ and $t = \vec{\theta}^\perp\cdot\vec{w}$, with the calculations in (3.5), we derive

$$\begin{aligned} f_\Omega(\vec{r}) &= \frac{1}{4\pi^2} \int_\Omega [\int_{R^2} \frac{\exp(-\mu\vec{v}\cdot(\vec{r}-\vec{w})e^{i\theta})}{\vec{\theta}\cdot(\vec{r}-\vec{w})} \vec{\theta}\cdot\nabla f(\vec{w})d\vec{w}]d\theta \\ &= \frac{1}{4\pi^2} \int_{R^2} f(\vec{w})d\vec{w} \int_\Omega [\vec{\theta}\cdot\nabla_{\vec{r}} \frac{\exp(-\mu\vec{v}\cdot(\vec{r}-\vec{w})e^{i\theta})}{\vec{\theta}\cdot(\vec{r}-\vec{w})}]d\theta. \end{aligned} \quad (4.3)$$

When $\Lambda^c = [0, 2\pi]$, the inner integral of (4.3) was proven to be the Dirac function. We study the following kernel function

$$K_\Omega(\vec{r}) = \frac{1}{4\pi^2}\int_\Omega \vec{\theta}\cdot\nabla_{\vec{r}}\frac{\exp(-\mu\vec{v}\cdot\vec{r}e^{i\theta})}{\vec{\theta}\cdot\vec{r}}d\theta. \tag{4.4}$$

For the subset $\Lambda$ in (4.1), rewrite equation (3.1) as follows:

$$\begin{aligned}f(\vec{r}) &= f_\Lambda(\vec{r}) + \int_{R^2} K_{\Lambda^c}(\vec{r}-\vec{w})f(\vec{w})d\vec{w} \\ &= 2f_\Lambda(\vec{r}) + \int_{R^2}[K_{\Lambda^c}(\vec{r}-\vec{w}) - K_\Lambda(\vec{r}-\vec{w})]f(\vec{w})d\vec{w}.\end{aligned} \tag{4.5}$$

Notice that $K_{\Omega+\pi}(\vec{r}) = K_\Omega(-\vec{r})$. By the constraint in (4.1), equation (4.5) becomes

$$\begin{aligned}f(\vec{r}) &= f_\Lambda(\vec{r}) + \int_{R^2} K_{\Lambda^c}(\vec{r}-\vec{w})f(\vec{w})d\vec{w} \\ &= 2f_\Lambda(\vec{r}) + \int_{R^2}[K_{\Lambda^c}(\vec{r}-\vec{w}) - K_{\Lambda^c}(\vec{w}-\vec{r})]f(\vec{w})d\vec{w}.\end{aligned} \tag{4.6}$$

If a unique solution of (4.6) exists, then the original function $f(x,y)$ can be recovered through solving the integral equation (4.6). So far, there have been two different algorithms for finding the solution of (4.6). One is using the first line of (4.6) and the other is based on the second line of (4.6). We shall introduce these two methods below.

According to [47], when $\mu$ is small, kernel function $K_{\Lambda^c}(\vec{r})$ defines a strictly contracting operator in $L^2(D)$. Then the first line of (4.6) alone is sufficient to deduce that there exists one and only one solution of (4.6). Let $K_\mu(\vec{r}) = K_{\Lambda^c}(\vec{r}) - K_{\Lambda^c}(-\vec{r})$, we observe that $K_\mu(\vec{r})$ is a skew-symmetric function. Restricted in $D$, $K_\mu(\vec{r})$ constructs a skew-symmetric bounded operator, denoted by $\mathbf{K}_\mu$, in $L^2(D)$. The simple fact of $||\mathbf{I} - \mathbf{K}_\mu||^2 = 1 + ||\mathbf{K}_\mu||^2$ implies that $(\mathbf{I} - \mathbf{K}_\mu)$ is coercive in $L^2(D)$ so that $(\mathbf{I} - \mathbf{K}_\mu)^{-1}$ exists in $L^2(D)$, here $\mathbf{I}$ stands for the identity operator. Now we conclude that there exists a unique solution of (4.6) and $f(x,y)$ satisfies the following integral equation

$$(\mathbf{I} - \mathbf{K}_\mu)f(x,y) = 2f_\Lambda(x,y). \tag{4.7}$$

So far, we have not considered the actual expressions of $K_{\Lambda^c}(\vec{r})$ and $K_\mu(\vec{r})$. In numerical realizations, a procedure has to be designed to calculate these kernel functions. Using the complex analysis inspired by [21], an explicit formula for $K_\mu(x,y)$ was derived in [36, 37]. For $\Lambda = [0, \pi]$, that formula takes the following expression

$$K_\mu(x,y) = \frac{\mu}{2\pi^2 x}[\frac{\sinh(\mu(y+ix))}{\mu(y+ix)} + \frac{\sinh(\mu(y-ix))}{\mu(y-ix)}]. \tag{4.8}$$

The author would like to comment that an alternate proof must exist to deduce (4.8) to avoid using the pseudo-math arguments based on the polar coordinate transform in $C^2$. At the moment, the author is trying to work out a different approach. A relaxed iterative method was developed in [36] to solve the integral equation (4.10). That iterative scheme reads as

$$f^{(n+1)}(\vec{r}) = 2\gamma f_\Lambda(\vec{r}) + (1-\gamma)f^{(n)}(\vec{r}) + \gamma\iint K_\mu(\vec{r}-\vec{w})f^{(n)}(\vec{w})d\vec{w}, \tag{4.9}$$

where $f^{(n)}(\vec{r})$ is the result from $n$-th iteration and $\gamma$ is the relaxation parameter that needs to be estimated using a prior information such as $\mu$ and the diameter of the support of $f(x,y)$. The implementation of (4.9) together with many simulations can be found in [36, 37]. Recently a more numerically efficient iterative scheme was developed in [47] without using the relaxation technique, thus there is no need to estimate $\gamma$. The iterative scheme of [47] is expressed as

$$f^{(n+1)}(\vec{r}) = f_\Lambda(\vec{r}) + \frac{1}{4\pi^2}\int_{\Lambda^c}e^{-\mu\vec{r}\cdot\vec{\theta}^\perp}\{\int_{-\infty}^\infty \frac{\cos(\mu(s-l))}{s-l}\frac{\partial}{\partial l}[\mathbf{R}_\mu f^{(n)}](\theta,l)dl\}|_{s=\vec{r}\cdot\vec{\theta}}\,d\theta. \tag{4.10}$$

The implementation of $\mathbf{R}_\mu f^{(n)}$ can be carried out by many fast projection generators developed for the iterative methods in the literature. It was proved in [47] that (4.10) converges when $\mu$ is small enough, in which the second term in the first line of (4.6) becomes a strictly contracting operator. Nonetheless, it remains open whether (4.10) converges for all $\mu$. Both (4.9) and (4.10) generated sound numerical reconstruction as reported in [36, 37, 47]. Numerically, kernel $K_{\Lambda^c}(\vec{r})$ can be computed ahead so that one may use the following iterative scheme

$$f^{(n+1)}(\vec{r}) = f_\Lambda(\vec{r}) + \iint K_{\Lambda^c}(\vec{r}-\vec{w}) f^{(n)}(\vec{w}) d\vec{w}. \tag{4.11}$$

As mentioned before, the DBH method intrinsically uses the projection data in 180º, so it readily provides an explicit procedure for the IERT with half-scan data for scanning ranges like $[0, \pi]$. However, the DBH method does not appear to be directly applicable to the noncontiguous subset $\Lambda$ defined in (4.1).

### B. Truncated-bin projection data

As shown in Fig 2, $p(\theta,s)$ may not be available outside of the thick circle. Prior to the DBH method, there is no any analytical reconstruction algorithm to accurately reconstruct $f(x,y)$ in a ROI based on partial measurements in the case of $\mu \neq 0$. In general, the global exact reconstruction from truncated-bin data is impossible as shown in [73] even the attenuation is not present. The DBH method converts a 2D reconstruction issue into a 1D inverse Hilbert transform, this ultimately leads to the possibility of accurately reconstructing $f(x,y)$ on some lines which fall in certain ROI. This method has been actively investigated for both SPECT and CT, see recent works for example [38-42, 78-84]. Compared with the FBP procedure, lately the DBH method does produce a lot of interests over other existing algorithms because of the idea of converting a 2D problem into a 1D problem.

Recall the data acquisition setting shown in Fig 2, the area inside of the black thick circle is detector's field of view (FOV). If the detector is not large enough, the FOV can not cover the entire object. However, if $f(x,y)$ has support in interval $[-L(y), L(y)]$ along that line, equation (3.24) reduces to

$$\hat{f}(x,y) = \int_{-L(y)}^{L(y)} \frac{\cosh(\mu(x-\tau))}{\pi(x-\tau)} f(\tau, y) d\tau. \tag{4.12}$$

By (2.11) and (4.12), we claim that $f(x,y)$ can be locally reconstructed in $[-L(y), L(y)]$ if $\hat{f}(x,y)$ can be calculated in the same interval. In other words, distribution $f(x,y)$ can be completely reconstructed by the DBH method between the two dashed lines for the data setting illustrated in Fig 2. In general, it is unclear whether the exact reconstruction still exists outside those two dashed lines. Many numerical simulation results have been reported in recent works [38-41] toward practical applications. Another benefit of the DBH method is that the DBP operation is adaptive to more complicated data acquisition geometries such as fan-beam or variable-focal-length fan-beam described in [25]. This makes it possible to exactly reconstruct fan-beam truncated-scan partial data [41].

In summary, with the better numerical stability and efficiency, the DBH method could have certain edge over the above mentioned iterative schemes based on (4.9) or (4.11). For practical applications, it is worth comparing the numerical behavior of those two types of methods besides their theoretical differences. However, one shortcoming of the DBH method is that the reconstruction in the non-exact region becomes much worse as revealed from the numerical simulations in many recent works for example [39-41]. For small truncation, from [[47]], the iterative scheme similar to (4.10) seems to yield better results compared with the DBH method.

### C. Limited-angle projection data

In history, limited-angle problem is a well-studied subject in the attenuation-free case, e.g., [85]. But it turns out to be quite complicated when the attenuation is present. Therefore, both practitioner and researchers try to avoid acquiring the limited-angle data in medical applications. In some industrial applications such as defects detection, engineers may only have the limited-angle data in locating the defects position inside the inspected materials. The study of limited-angle data reconstruction is still important.

Theoretically, the uniqueness of the reconstruction from limited-angle data is valid even for the attenuated Radon transform from [30, 34]. To author's knowledge, no stable numerical solution has ever existed in the literature. The iterative scheme (4.10) proposed in [47] for the inverse attenuated Radon transform is applicable to the IERT with limited-angle data. But the numerical results still remain unsatisfactory. I like to remark that the SVD method described in [62] for the limited-angle data reconstruction is very likely to exist for IERT as well.

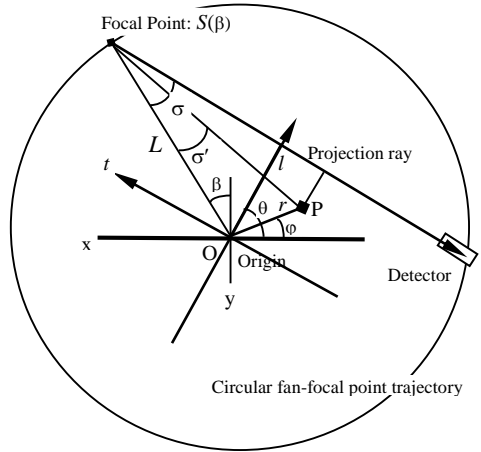

Fig 3: Each projection ray is uniquely determined by $(\beta, \sigma)$. Points O and S stand for the coordinate origin and one fan-focal point, respectively. Point P is an arbitrary reconstruction point. Here $\sigma$ denotes the angle between OS and the projection ray, $\sigma'$ denotes the angle between OS and PS, $\beta$ denotes the angle between OS and the y-axis, and $R$ is the radius of the circular orbit.

## V. Fan- and Cone-beam Data Reconstruction

In order to speed up the data acquisition and make full use of detection devices, the fan-beam or cone-beam formation has been widely chosen in commercial products. Thus, from the engineering point of view, the reconstruction from fan- or cone-beam data becomes more practical. Many analytical reconstruction methods for fan-beam geometry with considering the uniform attenuation correction had been reported in the literature, for example, [25, 48-50]. The more complicated reconstruction methods for 3D parallel- and cone-beam data acquisition geometries were also investigated in works such as [24, 51-55]. For the fan-beam geometry, most of the analytical algorithms in section III have the counterparts. The reconstruction algorithms for cone-beam geometry usually are computationally more intensive, thus approximate methods are preferred based on fan-beam reconstruction algorithms. The most popular method was [86]. Some extensions by my collaborators to SPECT imaging were made in [55, 87]. We will introduce the detailed fan- and cone-bin data formation and the geometric settings, and then provide a review on several analytical reconstruction algorithms.

### A. Fan-beam data acquisition geometry

A typical fan-beam data acquisition geometry with a circular fan-focal-point trajectory is shown in Fig 3, where projections can be represented by $(\beta, \sigma)$ for the arrowed line from a fan-focal point $S(\beta)$ with a ray angle $\sigma$. In this paper, the fan-beam uniformly attenuated projection $D_\mu f$ of function $f(x, y)$ is defined as

$$[D_\mu f](\beta, \sigma) = \int_0^\infty f(S(\beta) + \tau \vec{\alpha}(\beta, \sigma)) e^{\mu \tau} d\tau \qquad (5.1)$$

where $\sigma \in [-\sigma_m, \sigma_m]$, here $\sigma_m \in (0, \pi/2)$ denotes the maximum fan-subtending angle, and $\vec{\alpha}(\beta, \sigma)$ is a unit vector in $R^2$ to represent the direction from the focal point to the collimation hole as shown in Fig 3. Let $L$ be the radius of the circular fan-focal-point trajectory. The coordinates in parallel- and fan-beam geometries are related by

$$l = L\sin\sigma, \quad \theta = \sigma + \beta. \tag{5.2}$$

The line $(\beta, \sigma)$ in the fan-beam coordinates is the same line $(\theta, l)$ defined by (5.2) in the parallel-beam coordinates. The distance between the fan-focal point and the $l$-axis is $L\cos\sigma$. Thus the definitions of (1.1) and (5.1) can be related by the following relation

$$p(\sigma + \beta, R\sin\sigma) = e^{-L\cos\sigma}[D_\mu f](\beta, \sigma). \tag{5.3}$$

We use $g(\beta, \sigma)$ to represent the modified projection of (5.3) in the fan-beam data formation. By the chain rule of derivatives, we have

$$\frac{\partial g}{\partial \beta}(\beta, \sigma) = \frac{\partial p}{\partial \theta}(\theta, l) \text{ and } \frac{\partial g}{\partial \sigma}(\beta, \sigma) = \frac{\partial p}{\partial \theta}(\theta, l) + L\cos\sigma \frac{\partial p}{\partial l}(\theta, l). \tag{5.4}$$

Then, we obtain

$$\frac{\partial p}{\partial l}(\theta, l) = \frac{1}{L\cos\sigma}[(\frac{\partial}{\partial \sigma} - \frac{\partial}{\partial \beta})g](\beta, \sigma) \Rightarrow \{\text{defined as } \hat{p}^{fan}(\beta, \sigma)\}. \tag{5.5}$$

Let $K(r, \varphi, \beta)$ be the distance of SP, $\sigma'(r, \varphi, \beta)$ be the angle between SO and SP. Denote by $\vec{\alpha}^\perp$ the unit vector parallel to the $t$-axis in Fig 3. According to simple trigonometric calculations in [74], one has the following three geometric identities.

$$K(r, \varphi, \beta) = \sqrt{r^2 + L^2 + 2rL\sin(\beta - \varphi)}, \tag{5.6}$$

$$\sigma'(r, \varphi, \beta) = \arctan\frac{r\cos(\beta - \varphi)}{L + r\sin(\beta - \varphi)}, \tag{5.7}$$

$$r\cos(\theta - \varphi) - l = K(r, \varphi, \beta)\sin[\sigma'(r, \varphi, \beta) - \sigma]. \tag{5.8}$$

Now we give a brief overview on the analytical methods in the literature. The first analytical method for fan-bin data seemed to be [63], in which formula (3.28) was used. A direct extension of (3.21) to fan-beam data was investigated in [25] through using the Fourier interpolation along the rotation angle. The Fourier method was also investigated in [53, 54] to the fan-beam and cone-beam data reconstruction. Recently, an FBP-type algorithm was derived in [50] based on the inversion formula in [30, 31]. I would like to point out that the algorithm in [88] also can be directly extended to the fan-bin data reconstruction.

We first discuss the extension of the Hilbert transform based method in [88] to the fan-bin data. Let $\hat{s} = \vec{r} \cdot \vec{\theta}$ and $\hat{t} = \vec{r} \cdot \vec{\theta}^\perp$, and rewrite (3.1) as

$$f(\vec{r}) = \frac{1}{4\pi^2}\int_0^{2\pi} e^{-\mu\hat{t}}\{\cos(\mu\hat{s})\int_{-\infty}^\infty \frac{p'(\theta, l)}{\hat{s} - l}\cos(\mu l)dl + \sin(\mu\hat{s})\int_{-\infty}^\infty \frac{p'(\theta, l)}{\hat{s} - l}\sin(\mu l)dl\}d\theta. \tag{5.9}$$

In the fan-beam coordinates, we observe that $\hat{s} = r\cos(\beta + \sigma'(r, \varphi, \beta) - \varphi)$ and $\hat{t} = r\sin(\beta + \sigma'(r, \varphi, \beta) - \varphi)$. Then we express (5.9) in the following FBP algorithm,

$$f(r, \varphi) = \frac{1}{4\pi^2}\int_0^{2\pi}\frac{e^{-\mu\hat{t}}\cos(\mu\hat{s})}{K(r, \varphi, \beta)}[\int_{-\sigma_m}^{\sigma_m}\frac{\hat{p}^{fan}(\beta, \sigma)L\cos\sigma\cos(\mu L\sin\sigma)}{\sin[\sigma'(r, \varphi, \beta) - \sigma]}d\sigma]d\beta$$
$$+ \frac{1}{4\pi^2}\int_0^{2\pi}\frac{e^{-\mu\hat{t}}\sin(\mu\hat{s})}{K(r, \varphi, \beta)}[\int_{-\sigma_m}^{\sigma_m}\frac{\hat{p}^{fan}(\beta, \sigma)L\cos\sigma\sin(\mu L\sin\sigma)}{\sin[\sigma'(r, \varphi, \beta) - \sigma]}d\sigma]d\beta. \tag{5.10}$$

The calculation of $\hat{p}^{fan}(\beta, \sigma)$ needs the derivatives with respect to variable $\beta$. In SPECT imaging, the sampling for $\beta$ is usually very sparse, thus the ramp-filter was used in [50] to avoid the numerical errors due to the sparse sampling. Next we introduce the reconstruction method in [50] which was originally derived to handle the non-uniform attenuation correction. Again, we rewrite (3.1) in the following form

$$f(\vec{r}) = \frac{1}{4\pi^2} \int_0^{2\pi} e^{-\mu \hat{a}} \{ \int_{-\infty}^{\infty} [\cos(\mu(s-l)) \frac{\partial}{\partial s} \frac{1}{s-l} + \frac{\mu \sin(\mu(l-s))}{s-l}] p(\theta,l) dl \}|_{s=\hat{s}} \, d\theta$$

$$= \frac{1}{4\pi^2} \int_0^{2\pi} e^{-\mu \hat{a}} \{ [\cos(\mu \hat{s}) \int_{-\infty}^{\infty} \frac{\partial}{\partial s} \frac{1}{s-l} \cos(\mu l) + \sin(\mu \hat{s}) \int_{-\infty}^{\infty} \frac{\partial}{\partial s} \frac{1}{s-l} \sin(\mu l)] p(\theta,l) dl \}|_{s=\hat{s}} \, d\theta \quad (5.11)$$

$$+ \frac{\mu}{4\pi^2} \int_0^{2\pi} e^{-\mu \hat{a}} \{ \cos(\mu \hat{s}) \int_{-\infty}^{\infty} \frac{1}{\hat{s}-l} \sin(\mu l) - \sin(\mu \hat{s}) \int_{-\infty}^{\infty} \frac{1}{\hat{s}-l} \cos(\mu l)] p(\theta,l) dl \} d\theta.$$

For simplicity, we use $p_c(\theta,l) = \cos(\mu l) p(\theta,l)$ and $p_s(\theta,l) = \sin(\mu l) p(\theta,l)$. Let $R_\mu^{\hat{\omega}}(l)$ the ramp-filter kernel function which will be defined in Section VII, we express (5.11) as

$$f(\vec{r}) = \frac{1}{4\pi^2} \int_0^{2\pi} e^{-\mu \hat{a}} [\cos(\mu \hat{s}) \int_{-\infty}^{\infty} R_\mu^{\hat{\omega}}(\hat{s}-l) p_c(\theta,l) dl + \sin(\mu \hat{s}) \int_{-\infty}^{\infty} R_\mu^{\hat{\omega}}(\hat{s}-l) p_s(\theta,l) dl] d\theta$$

$$+ \frac{\mu}{4\pi^2} \int_0^{2\pi} e^{-\mu \hat{a}} [\cos(\mu \hat{s}) \int_{-\infty}^{\infty} \frac{p_s(\theta,l)}{\hat{s}-l} dl - \sin(\mu \hat{s}) \int_{-\infty}^{\infty} \frac{p_c(\theta,l)}{\hat{s}-l} dl] d\theta. \quad (5.12)$$

Define

$$p_c^{fan}(\beta,\sigma) = L\cos\sigma \cos(\mu L \sin\sigma) p^{fan}(\beta,\sigma), \quad (5.13)$$

$$p_s^{fan}(\beta,\sigma) = L\cos\sigma \sin(\mu L \sin\sigma) p^{fan}(\beta,\sigma). \quad (5.14)$$

Using the coordinate transformation, we derive

$$f(\vec{r}) = \frac{1}{4\pi^2} \int_0^{2\pi} \frac{e^{-\mu \hat{a}}}{K^2} [\cos(\mu \hat{s}) \int_{-\sigma_m}^{\sigma_m} R_\mu^{\hat{\omega}}(\sin(\sigma'-\sigma)) p_c^{fan}(\beta,\sigma) d\sigma + \sin(\mu \hat{s}) \int_{-\sigma_m}^{\sigma_m} R_\mu^{\hat{\omega}}(\sin(\sigma'-\sigma)) p_s^{fan}(\beta,\sigma) d\sigma] d\beta$$

$$+ \frac{\mu}{4\pi^2} \int_0^{2\pi} \frac{e^{-\mu \hat{a}}}{K} [\cos(\mu \hat{s}) \int_{-\sigma_m}^{\sigma_m} \frac{p_s^{fan}(\beta,\sigma)}{\sin(\sigma'-\sigma)} d\sigma - \sin(\mu \hat{s}) \int_{-\sigma_m}^{\sigma_m} \frac{p_c^{fan}(\beta,\sigma)}{\sin(\sigma'-\sigma)} d\sigma] d\beta. \quad (5.15)$$

Now it is evident that (5.15) does not include the partial derivative. The numerical simulations in [50] did produce reasonable reconstructions under the practical sampling grid of 128x128.

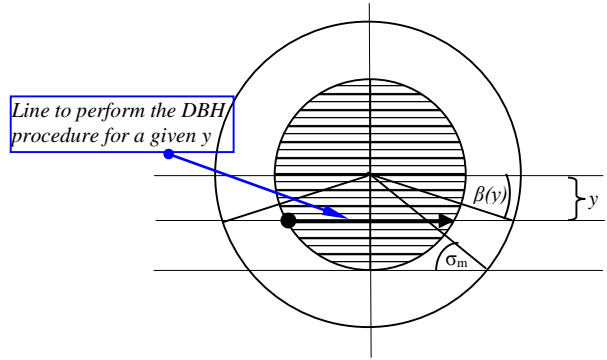

Fig 4: Horizontal lines in the inner unit disk for performing the DBH procedure in the fan-beam short-scan.

The application of the DBH method to the fan-beam data reconstruction has been investigated in our recent work [39, 41], where the derivation of the algorithm is rather straightforward through using (5.5). Let $\beta(y) = \arcsin(y/R)$ for $y \in (-1, 1)$ as illustrated in Fig 4. We transform the integration on $\theta$ in the parallel-beam geometry to the integration on $\beta$ in the fan-bema geometry. The calculation of $\hat{f}(x,y)$ in (3.24) can be carried out by the following formulas

$$\hat{f}(x,y) = -\frac{1}{2} \int_{\beta(y)-\pi/2}^{\pi/2-\beta(y)} e^{-\mu \hat{a}} \{ \frac{1}{K(r,\varphi,\beta)} [(\frac{\partial}{\partial \sigma} - \frac{\partial}{\partial \beta}) p^{fan}](\beta,\sigma'(r,\varphi,\beta)) \} d\beta, \quad (5.16)$$

Similarly, in order to avoid the derivatives with respect to variable $\beta$, one may rewrite (5.16) as follows

$$\hat{f}(x,y) = -\frac{1}{2}\int_{\beta(y)-\pi/2}^{\pi/2-\beta(y)} \frac{e^{-\mu\hat{a}}}{K(r,\varphi,\beta)}[\frac{\partial}{\partial\sigma}p^{fan}](\beta,\sigma'(r,\varphi,\beta))d\beta \qquad (5.17)$$

$$+\frac{1}{2}[\frac{e^{-\mu\hat{a}}}{K(r,\varphi,\beta)}p^{fan}(\beta,\sigma'(r,\varphi,\beta))]\Big|_{\beta=\beta(y)-\pi/2}^{\beta=\pi/2-\beta(y)} - \int_{\beta(y)-\pi/2}^{\pi/2-\beta(y)} p^{fan}(\beta,\sigma'(r,\varphi,\beta))d\frac{e^{-\mu\hat{a}}}{K(r,\varphi,\beta)}.$$

It is also worthwhile to mention that the weight $1/K(r,\varphi,\beta)$ may cause singularity in using (5.17) when $r \to R$, see detailed analysis on the effect of that weight in [88]. It was proved in the appendix of [88] that $\theta$ and $\beta$ satisfy the following differential relation:

$$d\theta = d(\beta + \sigma'(r,\varphi,\beta)) = \frac{R\cos\sigma'(r,\varphi,\beta)}{K(r,\varphi,\beta)}d\beta. \qquad (5.18)$$

We express $\sigma'(r,\varphi,\beta)$ in the parallel-beam coordinate as $\hat{\sigma}(r,\varphi,\theta)$, i.e.,

$$\hat{\sigma}(r,\varphi,\theta) = \arcsin\frac{\vec{r}\cdot\vec{\theta}}{R} = \arcsin\frac{r\cos(\theta-\varphi)}{R}. \qquad (5.19)$$

The first equality in (5.19) is due to (5.2). In order to remove the factor $1/K(r,\varphi,\beta)$, by using (5.18) and (5.19) we may rewrite (5.16) in the following expression

$$\hat{f}(x,y) = -\frac{1}{2\pi}\int_{-\pi/2}^{\pi/2} e^{\mu r\sin(\theta-\varphi)}\hat{p}^{fan}(\theta-\sigma'(r,\varphi,\theta),\sigma'(r,\varphi,\theta))d\theta. \qquad (5.20)$$

The reconstruction of $f(x,y)$ from $\hat{f}(x,y)$ shall follow the same procedure as detailed for the parallel-beam data reconstruction in Section 3.

The CHE method for the fan-beam-like data can be derived from (3.21) as follows,

$$f(r,\varphi) = \frac{1}{\pi}\int_0^{2\pi} e^{\mu r\sin(\theta-\varphi)}d\theta \int_{-\infty}^{\infty} R_\mu^{\hat{\omega}}(r\cos(\theta-\varphi)-l)\sum_n p_n(l)e^{in\theta}dl$$

$$= \frac{L}{2\pi}\sum_n e^{in\varphi}\int_{-\sigma_m}^{\sigma_m} p_n(L\sin\sigma)\cos\sigma d\sigma \int_0^{2\pi} R_\mu^{\hat{\omega}}(r\cos\theta-L\sin\sigma)e^{\mu r\sin\theta}e^{in\theta}d\theta. \qquad (5.21)$$

Using the Fourier series expansion based interpolation suggested in [25], coefficients $p_n(L\sin\sigma)$ can be calculated from $p^{fan}(\beta,\sigma)$ through the following phase-shift formula,

$$p_n(L\sin\sigma) = \int_0^{2\pi} p(\theta,L\sin\sigma)e^{-in\theta}d\theta = e^{-in\sigma}\int_0^{2\pi} p^{fan}(\beta,\sigma)e^{-in\beta}d\beta = e^{-in\sigma}p_n^{fan}(\sigma). \qquad (5.22)$$

where $p_n^{fan}(\sigma)$ is the Fourier series coefficients of $p^{fan}(\beta,\sigma)$ with respect to the first variable. And integral $\int_0^{2\pi} R_\mu^{\hat{\omega}}(r\cos\theta-L\sin\sigma)e^{\mu r\sin\theta}e^{in\theta}d\theta$ of (5.16) can be calculated by the fast FT. With the advent of the FBP algorithm, the CHE may not be preferred for fan-bin data reconstruction since it requires special data sampling grid in order to use the fast FT.

In the fan-beam coordinates, (3.29) becomes

$$f(0) = \int_0^{\sigma_m} \frac{\cos(\mu L\sin\sigma)\cos\sigma}{\pi\sin\sigma}\{\frac{1}{4\pi}\int_0^{2\pi}[\hat{p}^{fan}(\theta-\sigma',-\sigma)-\hat{p}^{fan}(\theta-\sigma',\sigma)]d\theta\}d\sigma. \qquad (5.23)$$

Inversion formula (5.23) was previously used in [48] for fan-beam data reconstruction, but it requires too much computation and thus may be obsolete.

**Remark 5**. The circular orbit of fan-focal points has been discussed so far. Actually, the FBP and DBH algorithms also can be easily extended to non-circular orbits such as the elliptic curve in [87].

### B. Cone-beam data acquisition geometry

The research of cone-bin data acquisition and reconstruction had been very active dated to 1980's. In order to maintain the mathematically exact reconstruction, the trajectory of the cone-focal points has to meet the sufficient condition stated in [89, 90]. As a result, the reconstruction method became much more complicated. Practically, the planar circle and spiral are two favorable trajectories from the system design and manufacture point of view. When the attenuation is not present, many exact reconstruction algorithms are based on two classical works [89, 90]. However, the actual data even from research systems never met the sufficient conditions required in many algorithms. For simplicity, in this paper, we only discuss the planar circular trajectory for the cone-focal points. So far, the FDK algorithm in [86] has been the most successful method from the practical point of view. That method is a direct extension of the fan-beam FBP algorithm to the cone-bean data. Due to its simplicity in data acquisition and the computational efficiency, the FDK algorithm had been extensively investigated for applications in both SPECT and CT in last two decades. Recently, numerical results of the FDK method to include the nonuniform attenuation correction were reported in [55]. In this paper, I shall sketch out the FDK procedure for the IERT based on the FBP algorithm (5.15).

The data formation in the circular trajectory used in FDK method is shown in Fig 5. Different from [86], we use the equiangular data sampling throughout this paper. All the results should be valid to the equally-spaced data formation too. On the detector plane, each projection ray $QS$ is represented by $(\beta, \sigma, \hat{z})$, the cone-beam exponential Radon transform, denoted by $[D_\mu f](\beta, \sigma, \hat{z})$, takes the following definition

$$[D_\mu f](\beta, \sigma, \hat{z}) = \int_0^\infty f(S(\beta) + \tau \vec{\alpha}(\beta, \sigma, \hat{z})) e^{\mu \tau} d\tau \tag{5.24}$$

where $\vec{\alpha}(\beta, \sigma, \hat{z})$ is the unit vector from $S$ to $Q$. Similar to the fan-beam FBP algorithm, the FDK method contains two steps: filtering on a set of space planes and backprojection on the projection rays passing through reconstruction point $P$, denoted by $\vec{u} = (x, y, z)$. The cylinder coordinate $\vec{u} = (r, \varphi, z)$ is frequently used. The detailed coordinates of the filtering planes are illustrated in Fig 5, where the cone-focal point and the detector plane constructs a unique pair. Let $R$ be the distance between the cone-focal point and the detector plane, and denote $L_{\hat{z}} = L/\lambda(\hat{z})$, here $\lambda(\hat{z}) = \cos(\arctan(\hat{z}/R))$ is the scaling factor between $xy$-plane and tilted plane $QST$. We define modified projection $p^{cb}(\beta, \sigma, \hat{z})$ as follows

$$p^{cb}(\beta, \sigma, \hat{z}) = e^{-\mu L_{\hat{z}} \cos\sigma} [D_\mu f](\beta, \sigma, \hat{z}). \tag{5.25}$$

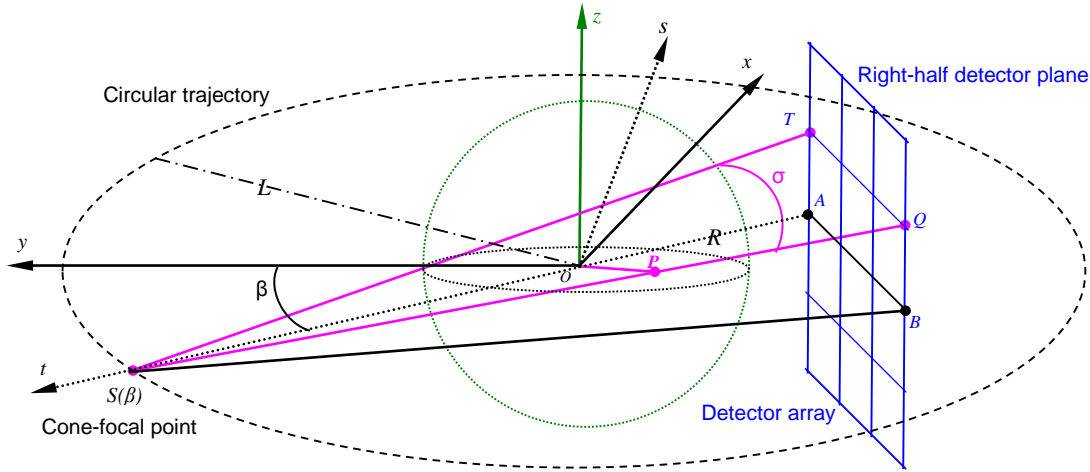

Fig 5: Let $o - (x, y, z)$ stand for the Cartesian coordinate in $R^3$, here $o$ is the origin. The circular trajectory locates on the $xy$-plane with radius $L$. $S$ and plane $ABQT$ is one pair of cone-focal point and detector array. $AB$ is the intersection line of detector plane and the circular trajectory plane. Assume $P$ at $(x, y, z)$ is the point at which the value needs to be reconstructed. Line $AS$ passes the origin and is perpendicular to the detector plane with length $R$, and projection ray $SP$ intersects with detector plane at $Q$. Notice only the right half detector array is shown here. Point $B$ is the projection of $Q$ on the $xy$-plane and line $TQ$ is parallel to line $AB$. The coordinate at $Q$ is expressed as $(\beta, \sigma, z)$, here $\beta$ is the angle between $OS$ and $y$-axis, $\sigma$ is the angle between $QS$ and $TS$, and $z$ is the distance of $QB$.

The filtering in the FDK method is performed on the plane *QST* while the backprojection needs to be performed on a set of lines which do not belong to one plane. Assume that $\vec{u} = (x, y, z)$ is the point at which the value needs to be reconstructed. We introduce several geometry parameters

$$\hat{z}(r,\varphi,z,\beta) = \frac{Rz}{L + r\sin(\beta - \varphi)}, \tag{5.26}$$

$$\hat{K}(r,\varphi,z,\beta) = \sqrt{z^2 + r^2 + L^2 + 2rL\sin(\beta - \varphi)}, \tag{5.27}$$

$$\hat{\sigma}(r,\varphi,z,\beta) = \arctan(\frac{\lambda(\hat{z}(r,\varphi,z,\beta))r\cos(\beta - \varphi)}{L + r\sin(\beta - \varphi)}), \tag{5.28}$$

where $\hat{z}(r,\varphi,z,\beta)$ is the distance of *QB*, $\hat{K}(r,\varphi,z,\beta)$ is the distance of *SP*, and $\hat{\sigma}(r,\varphi,z,\beta)$ is the angle between *PS* and *AS*.

Due to the sparse sampling along the trajectory, we try to avoid the partial derivatives in 3D space. Thus, we only discuss the extension of (5.15) following the FDK procedure. Define the weighted projections

$$p_c^{cb}(\beta,\sigma,\hat{z}) = L_{\hat{z}} \cos\sigma \cos(\mu L_{\hat{z}} \sin\sigma) p^{cb}(\beta,\sigma,\hat{z}), \tag{5.29}$$

$$p_s^{cb}(\beta,\sigma,\hat{z}) = L_{\hat{z}} \cos\sigma \sin(\mu L_{\hat{z}} \sin\sigma) p^{cb}(\beta,\sigma,\hat{z}). \tag{5.30}$$

The ramp-filtered and Hilbert-transformed data are

$$\hat{p}_c^{cb}(\beta,\sigma,\hat{z}) = \int_{-\sigma_m}^{\sigma_m} R_\mu^{\hat{\omega}}(\sin(\sigma - \tau)) p_c^{cb}(\beta,\tau,\hat{z}) d\tau, \tag{5.31}$$

$$\hat{p}_s^{cb}(\beta,\sigma,\hat{z}) = \int_{-\sigma_m}^{\sigma_m} R_\mu^{\hat{\omega}}(\sin(\sigma - \tau)) p_s^{cb}(\beta,\tau,\hat{z}) d\tau, \tag{5.32}$$

$$\tilde{p}_c^{cb}(\beta,\sigma,\hat{z}) = \int_{-\sigma_m}^{\sigma_m} p_c^{cb}(\beta,\tau,\hat{z})/\sin(\sigma - \tau) d\tau, \tag{5.33}$$

$$\tilde{p}_s^{cb}(\beta,\sigma,\hat{z}) = \int_{-\sigma_m}^{\sigma_m} p_s^{cb}(\beta,\tau,\hat{z})/\sin(\sigma - \tau) d\tau. \tag{5.34}$$

For the sake of simplicity, we use $\hat{z}$ for $\hat{z}(r,\varphi,z,\beta)$ and $\hat{\sigma}$ for $\hat{\sigma}(r,\varphi,z,\beta)$, respectively. With the above notation, the FDK formula is expressed as

$$f(r,\varphi,z) = \frac{1}{4\pi^2} \int_0^{2\pi} \frac{e^{-\mu\hat{t}}}{\hat{K}^2(r,\varphi,z,\beta)} [\cos(\mu\hat{s})\hat{p}_c^{cb}(\beta,\hat{\sigma},\hat{z}) + \sin(\mu\hat{s})\hat{p}_s^{cb}(\beta,\hat{\sigma},\hat{z})] d\beta$$
$$+ \frac{\mu}{4\pi^2} \int_0^{2\pi} \frac{e^{-\mu\hat{t}}}{\hat{K}(r,\varphi,z,\beta)} [\cos(\mu\hat{s})\tilde{p}_s^{cb}(\beta,\hat{\sigma},\hat{z}) - \sin(\mu\hat{s})\tilde{p}_c^{cb}(\beta,\hat{\sigma},\hat{z})] d\beta \tag{5.35}$$

where

$$\hat{s} = r\cos(\lambda(\hat{z})(\beta - \varphi) + \hat{\sigma}(r,\varphi,z,\beta))/\lambda(\hat{z}), \tag{5.36}$$

$$\hat{t} = r\sin(\lambda(\hat{z})(\beta - \varphi) + \hat{\sigma}(r,\varphi,z,\beta)]/\lambda(\hat{z}). \tag{5.37}$$

Since $\beta - \varphi$ and $\hat{\sigma}(r,\varphi,z,\beta)$ are measured on different planes, both $r$ and $\beta - \varphi$ need to be scaled to the measurement on the plane *QST*. We want to mention that $\lambda(\hat{z})(\beta - \varphi)$ is still approximate and its impact on the reconstruction has not been investigated. For $\mu = 0$, formula (5.33) becomes the conventional FDK method for the equiangular projection data.

## VI. Noise Dilemma

For simplicity, we only consider the noise effects of the FBP algorithm for the parallel-beam projection data. From now on, we should understand the projection data in a set of discrete measurements as follows

$$\tilde{p}(\theta_k, s_m) = p(\theta_k, s_m) + n(\theta_k, s_m) \tag{6.1}$$

where $1 \leq k \leq K$ and $1 \leq m \leq M$, $p(\theta_k, s_m)$ is the noise-free ERT defined by (1.1), and $n(\theta_k, s_m)$ is the error caused by either the imperfect data acquisition devices or the statistic nature of the data. Certainly, both $p(\theta_k, s_m)$ and $n(\theta_k, s_m)$ are unknown for a given measurement $\tilde{p}(\theta_k, s_m)$. Because of the Poisson nature of the photon counts, a common assumption is that $\tilde{p}(\theta_k, s_m)$ is a Poisson random variable with mean value of $p(\theta_k, s_m)$ and $\tilde{p}(\theta_k, s_m)$ is independent to each other.

The first dilemma is on whether the aforementioned independency assumption holds in a rigorous sense. In author's view, that assumption looks quite dubious. For the sake of simplicity, I assume that the attenuation is not present. From [71] and the nature of the photon counts, it is reasonable to assume that the discretized $f(x_i, y_j)$ can be regarded as a set of independent Poisson variants. It follows that measurements $\tilde{p}(\theta_k, s_m)$ and $\tilde{p}(\theta_k + \pi, -s_m)$ are the summation of the same Poisson variants, thus they have exactly the same statistics, and certainly are not independent to each other. This suggests that the common assumption about the independency condition may not be true. Anyway, all the statistic analysis needs the independency assumption.

With the independency assumption, one interesting result from [58] states that the noise level can be estimated based on the measurement $\tilde{p}(\theta_k, s_m)$. To author's knowledge, this has been the only result that can be mathematically well established. We define the noise percentage of $\tilde{p}(\theta_k, s_m)$ by

$$err = \sqrt{\sum_{k,m} n^2(\theta_k, s_m) / \sum_{k,m} p^2(\theta_k, s_m)} . \tag{6.2}$$

As aforementioned, $err$ is not immediately available because $p(\theta_k, s_m)$ and $n(\theta_k, s_m)$ cannot be directly measured. Recently, a formula of estimating $err$ based on measurements $\tilde{p}(\theta_k, s_m)$ was given in [58]. That formula reads as

$$err \approx \sqrt{\sum_{k,m} \tilde{p}(\theta_k, s_m) / (\sum_{k,m} \tilde{p}^2(\theta_k, s_m) - \sum_{k,m} \tilde{p}(\theta_k, s_m))} . \tag{6.3}$$

Formula (6.3) was derived through Chebyshev's inequality, and holds true with a probability close to 1 as proved in [58]. It is very amazed that (6.3) gives a very high accuracy in numerical simulations.

The conventional wisdom is that the noise can be suppressed through filtering. This may be true in many applications that involve signal processing. The second dilemma is that this common strategy does not necessarily hold in the image reconstruction. We express (6.1) in its continuous form as

$$\tilde{p}(\theta, s) = p(\theta, s) + n(\theta, s) . \tag{6.4}$$

Denote by $\mathbf{R}_\mu^{-1}$ the IERT, and define the following two functions,

$$n_R(\theta, s) = \mathbf{R}_\mu \mathbf{R}_\mu^{-1} n(\theta, s), \quad n_N(\theta, s) = n(\theta, s) - n_R(\theta, s). \tag{6.5}$$

Because of the equality $\mathbf{R}_\mu \mathbf{R}_\mu^{-1} p(\theta, s) = p(\theta, s)$, we derive

$$n_N(\theta, s) = \tilde{p}(\theta, s) - \mathbf{R}_\mu \mathbf{R}_\mu^{-1} \tilde{p}(\theta, s) . \tag{6.6}$$

Notice that $n_N(\theta, s)$ can be estimated by $\tilde{p}(\theta, s)$ because of identity $\mathbf{R}_\mu^{-1} n_N(\theta, s) \equiv 0$. This implies that the total noise $n(\theta, s)$ can be decomposed into two components: $n_N(\theta, s)$ is called the null space noise because it annihilates in the IERT, and $n_R(\theta, s)$ is called the range space noise since it meets the range condition. Obviously, there is no need to filter $n_N(\theta, s)$ since it has been filtered out in the reconstruction. This suggests that it only makes sense to filter $n_R(\theta, s)$ in order to reduce the noise in the reconstructed image. Is there any criterion that can ensure the filtering process does reduce $n_R(\theta, s)$. No conclusion has been drawn on this question to author's knowledge. The much worse observation is that the null space noise $n_N(\theta, s)$ was the dominant component in the total noise $n(\theta, s)$ as reported in [59]. The implication of this observation is that filtering projection noise is by large a blind process without knowing whether the noise in the reconstructed image can be mitigated. One extreme example is that a filter that happens to completely remove $n_N(\theta, s)$ while keeping $n_R(\theta, s)$ unchanged does not help mitigate the reconstruction noise at all although it may dramatically reduce the error percentage of the projection data. As a result, the variance analysis itself is incomplete to understand the noise propagation in the image reconstruction.

Well, it remains unclear about whether $n_R(\theta, s)$ can be filtered, but the experience is that filters do make images look visually smooth at least. The consensus is to reduce $n(\theta, s)$ regardless of whether it helps filter the reconstruction noise. Let $\tilde{f}(x_i, y_j)$ be the reconstructed image from the noise data $\tilde{p}(\theta_k, s_m)$, and $\bar{f}(x_i, y_j)$ is the reconstructed image from filtered projection data $\bar{p}(\theta_k, s_m)$. The ideal filter should decrease the error percentage in the reconstructed image, i.e.,

$$\sum_{i,j}(\bar{f}(x_i, y_j) - f(x_i, y_j))^2 < \sum_{i,j}(\tilde{f}(x_i, y_j) - f(x_i, y_j))^2. \tag{6.7}$$

It seems that there are no any filtering methods in the literature that have been proven to meet (6.7) in a convincing manner. The alternate way is to find other criterion to justify the use of filters to reduce the reconstruction noise. The concept of variance comes into play in many investigations on the noise analysis and treatment since virtually all filters could reduce the variance to some extent, for example the variance formulas of several filters in [77]. But reducing the variance of the reconstructed image is irrelevant to meeting (6.7).

Stochastic process is a quite profound subject in mathematics, and sometimes is used in analyzing the noise propagation from the projection to the reconstructed image. In this method, $\tilde{p}(\theta, s)$ is assumed to be a random process, denoted by $\mathbf{p}(\theta, s)$. The IERT can be regarded as a linear transform on $\mathbf{p}(\theta, s)$. The difficulty of such analysis for the IERT is that $\mathbf{p}(\theta, s)$ is not even a weakly stationary process and the IERT is not a linear operator in the same function space (recall that it is a mapping from functions on $S^1 \times I$ to functions on $R^2$). This implies that many existing results for the random process transformation are not available to the IERT in a rigorous sense. As a result, the common criterion of minimizing the variance of the reconstructed image lacks a direct connection to estimating the mean of the reconstructed image.

In general, "filtering noise" may be an engineering activity without needing a well-defined discipline, especially sometimes it is very subjective to define a criterion to measure the level of noise reduction by a specific filtering method in using (3.1). For example, it was proved in [56, 57] long time ago that the FBP algorithm is nearly optimal when the attenuation is not present. For the IERT, it was proved in [22] that weighting the two redundant terms in (3.11) may reduce the covariance of the projection data in the frequency domain. A natural question arises whether a universal criterion can be used to justify the "optimality" in those works. On the other hand, complicated interpolations are needed to handle the nonlinearity in the implementation of the Fourier method (3.11), which could affect the numerical results to a level that may not be ignored in the optimality analysis. In the theory of random process [71], given a random process, the statistics such as the mean and variance should be determined. Thus, minimizing the variance of a random process does not directly relate to finding the mean of that random process.

In conclusion, this author's view is that smoothing $\tilde{p}(\theta, s)$ is more like engineering experiences for improving the visual perception instead of actually reducing the noise. One interesting comment in [77] is that the noise treatment is more like an art than the science. I guess this is true in the research community.

## VII. Computer Simulation

We have seen many different inversion formulas for the IERT. These inversion formulas actually indeed reveal many different characteristics in terms of the numerical stability and robustness to noise contamination. In this paper we provide computer simulation results for several classical FBP algorithms and the recent DBH algorithm for parallel-beam full- and half-scan data. In particular, the reconstruction from noisy data should provide readers a glimpse of the noise dilemma and indicate the importance of noise treatment. For practitioners of various reconstruction methods in engineering developments and research study, we provide the demo software to exercise many algorithms reviewed in this paper. The software can be downloaded from www.cubic-imaging.com.

## A. Numerical preliminary

Many digital filters have been specifically designed for the image reconstruction algorithms in [72, 77]. In this paper several popular filters will be reviewed. We use $\Delta$ for the sampling interval of distance variables and $\Phi$ for the sampling interval of angular variables. Let $\varpi$ be the cut-off frequency in the design of digital filters. The ramp convolution kernel $R_{rec}^{\varpi}(l)$ is defined by

$$R_{rec}^{\varpi}(l) = \frac{1}{2\pi}\int_{\varpi>|\omega|}|\omega|e^{i\omega l}d\omega = \frac{\varpi\sin(l\varpi)}{\pi l} + \frac{\cos(l\varpi)-1}{\pi l^2}. \tag{7.1}$$

The Shepp-Logan filter $R_{SL}^{\varpi}(l)$ is defined by

$$\begin{aligned}R_{SL}^{\varpi}(l) &= \frac{1}{2\pi}\int_{\varpi>|\omega|}|\omega|\frac{2\varpi\sin(0.5\pi\omega/\varpi)}{\pi\omega}e^{i\omega l}d\omega \\ &= \frac{\varpi}{\pi^2}[\frac{1}{0.5\pi/\varpi+l}(1+\sin(l\varpi)) + \frac{1}{0.5\pi/\varpi-l}(1-\sin(l\varpi))]\end{aligned} \tag{7.2}$$

The Hilbert transform kernel function, denoted by $H^{\varepsilon}(l)$ for small $\varepsilon>0$, takes the following form

$$H^{\varepsilon}(l) = \frac{l}{\pi(l^2+\varepsilon^2)}. \tag{7.3}$$

We mention two other Hilbert transform kernels used in the computer simulations of [31, 32] for the inverse attenuated Radon transform.

$$H_{rec}^{\varpi}(l) = -\frac{i}{2\pi}\int_{\varpi>|\omega|}\text{sign}(\omega)e^{il\omega}d\omega = \frac{1}{\pi l}(1-\cos(l\varpi)). \tag{7.4}$$

$$H_{KL}(n) = \begin{cases} 2n\ln 2/\pi & n=0,1 \\ \frac{1}{\pi}(\frac{1}{n}\ln\frac{n^2-1}{n^2}+\ln\frac{n+1}{n-1}) & n>1. \end{cases} \tag{7.5}$$

For simplicity, the partial derivative is implemented as the central difference and the function evaluation takes the linear or bilinear interpolation depending on the dimension of the considered variables.

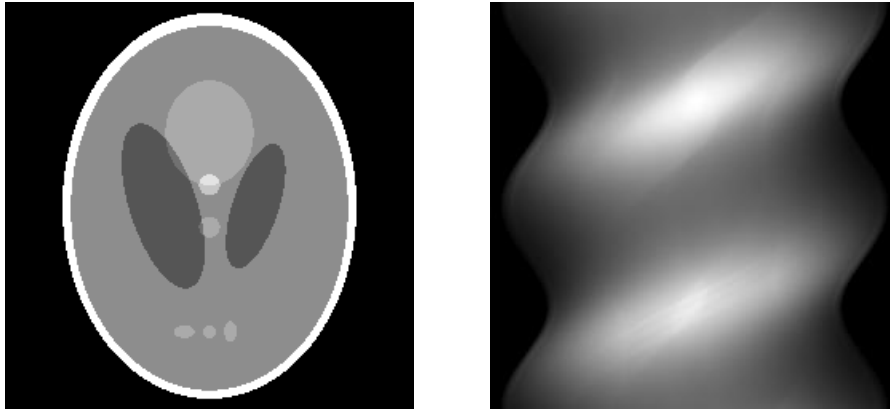

Fig 6: Left is the Shepp-Logan phantom and right is the accurately computed projection data.

The density function $f(x,y)$ takes the popular Shepp-Logan phantom from [12] after minor changes to its intensity values. All used projection data $p(\theta,l)$ are analytically calculated based on the Shepp-Logan phantom, thus they are mathematically exact without discrete errors. The Shepp-Logan phantom and the generated projection data are shown in Fig 6. Theoretically, to meet the Shannon sampling condition in the reconstructed image, the angular sampling and distance sampling of $p(\theta,l)$ needs satisfy $\pi\Phi=\Delta$ according to [73]. However, this condition is very unrealistic in the clinic SPECT. In this paper, projections $p(\theta,l)$ are evenly sampled over $[0,2\pi]\times(-1,1)$ in a 256x256 grid. In the simulation of noisy projection, the Poisson generator in [76] is applied to generate the photon counts. Function $f(x,y)$ will be reconstructed over $(-1,1)\times(-1,1)$ in a 256x256 grid. Without specifically stating, in all simulations we choose $\mu=3.0$, which is equivalent to the linear attenuation coefficient of $0.2\text{cm}^{-1}$ in brain SPECT with the brain matter of diameter 30cm. In summary, in all numerical computations we use

$$\Delta = 1/128 \text{ and } \Phi = \pi/128. \tag{7.6}$$

Under these sampling assumptions, the cut-off frequency is $\varpi = \pi/\Delta$ and the sampling interval in the frequency domain should be 0.5.

## B. Simulation results by (3.1)

Virtually, the numerical implementation of (3.1) follows the same procedure as the FBP algorithm in [12] except with a modified filter to exclude the low frequency component and a weighted back-projection. Define $R_\mu^{\hat{\omega}}(l) = R_{SL}^{\hat{\omega}}(l) - R_{rec}^\mu(l)$, $\hat{\omega} = \pi/\Delta$ is the cut-off frequency, we discretize the FBP algorithm of [13] as follows

$$f(r,\varphi) = \frac{1}{2\pi}\int_0^{2\pi} e^{\mu r \sin(\theta-\varphi)}d\theta \int_{-\infty}^{\infty} R_\mu^{\hat{\omega}}(r\cos(\theta-\varphi)-l)p(\theta,l)dl. \tag{7.7}$$

The numerical implementation of (7.7) contains two steps.

1. Filtering projection data: $q(\theta,s) = \int_{-\infty}^{\infty} R_\mu^{\hat{\omega}}(s-l)p(\theta,l)dl.$ (7.8)

2. Backprojecting the filtered data: $f(r,\varphi) = \frac{1}{2\pi}\int_0^{2\pi} e^{\mu r \sin(\theta-\varphi)} q(\theta, r\cos(\theta-\varphi))d\theta.$ (7.8)

The only difference between (7.8-7.9) and the implementation from [31, 32] is in that the ramp filtering is decomposed into two steps of the numerical difference and the weighted Hilbert transform. That is to say, for $\hat{\varepsilon} = \Delta/8$ and $H_\mu^{\hat{\varepsilon}}(l) = \cos(\mu l) H^{\hat{\varepsilon}}(l)$, the filtering process in [31, 32] is equivalent to

$$q(\theta,s) = \int_{-\infty}^{\infty} H_\mu^\varepsilon(s-l) \frac{p(\theta,l+\Delta)-p(\theta,l-\Delta)}{2\Delta} dl. \tag{7.10}$$

The reconstructed images are shown in Fig 7. Notice that the reconstructed image using filter (7.8) looks sharp but includes more strike artifacts while the reconstructed image using (7.10) seems to be smooth with reduced artifacts. This is one example of using multiple implementations of the same inversion formula to smooth the reconstructed image. Indeed, different implementations yield slightly different numerical results, at least from the visual perception.

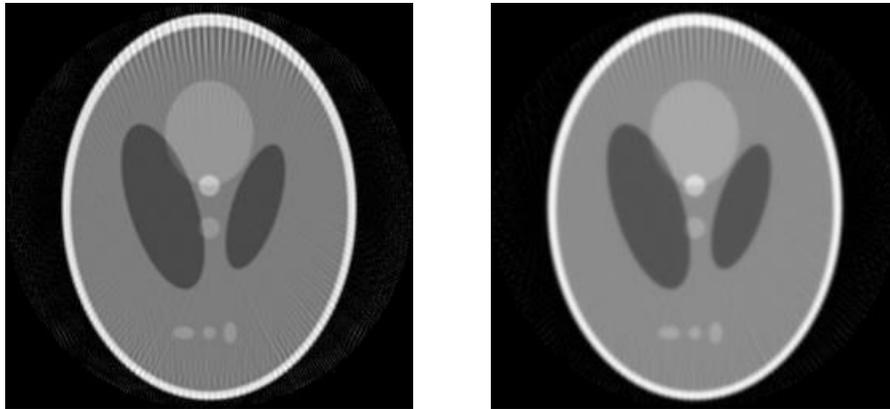

Fig 7: Left is the reconstructed image using (7.7) and right is the reconstructed image using (7.10).

To further explore the cause of aforementioned different reconstruction results, we notice that

$$\frac{p(\theta,l+\Delta)-p(\theta,l-\Delta)}{2\Delta} = \frac{1}{2\pi}\int_{-\infty}^{\infty} \tilde{p}(\theta,\omega)\omega e^{is\omega}[\frac{\sin(\Delta\omega/2)}{\Delta\omega/2}]d\omega. \tag{7.11}$$

Factor $\sin(\Delta\omega/2)/(\Delta\omega/2)$ is an intrinsic filter in use of the central difference. More suppression of the high frequency component, smoother the reconstructed image would become. We believe that above reconstruction differences come from the numerical behavior of those two digital filters.

### C. Simulation procedure by (3.11) and (3.18)

The implementation of (3.11) requires that the sampling rate for both the angular variable and the distance variable should be the power of 2 in order to use the FFT. For the same reason the sampling rate of the angular variable in realizing (3.21) needs to be the power of 2 as well. Theoretically, formulas (3.11) and (3.21) are just two different expressions of (3.11) by using the FT and FSE. Thus numerically it should not be generating much difference among these implementations as shown in many existing works for example [20-23, 25].

The calculation of $\tilde{p}_n(\omega)$ from $p(\theta,l)$ is rather straightforward through the FT and FSE. The key issue comes from the polar coordinate expression $\tilde{f}(\rho,\psi)$ since the inverse FFT (IFFT) is not available to derive $f(x,y)$ from $\tilde{f}(\rho,\psi)$. More precisely speaking, relation (3.11) converts the attenuated projection to the FT of $f(x,y)$ in the polar coordinates system. The subsequent step has to recover $f(x,y)$ in the Cartesian coordinate from $\tilde{f}(\rho,\psi)$. If $p(\theta,l)$ is evenly sampled with respect to $l$, functions $\tilde{f}_n(\rho)$ and $\tilde{f}(\rho,\psi)$ become irregularly sampled on $\rho$. Thus the interpolation to handle the nonlinearity in $\sqrt{\omega^2-\mu^2}$ can not be avoided regardless of the reconstruction method. Based on comments in [77], the interpolation of $\tilde{f}(\rho,\psi)$ to $\tilde{f}(\xi_x,\xi_y)$ is not a good choice though it was either used or discussed in works for example [17, 21-23], in particular [23] provides a summary on these methods. Here I will provide a slightly different version of the algorithm in [20] to avoid the direct interpolation through (3.9) to calculate $f_n(r)$ by using the numerical integral on an irregular sampling grid. The detailed numerical procedure contains three steps.

1. Compute $\tilde{p}_n(\omega)$ from $p(\theta,l)$ by the IFFT. (7.12)

2. Compute $\tilde{f}_n(\sqrt{(0.5m)^2-\mu^2})$ for $m \geq 2|\mu|$, $m$ is positive integers. (7.13)

3. Compute $J_n(k\Delta\sqrt{(0.5m)^2-\mu^2})$ from (2.3) by FFT for integers $m$ and $k$. (7.14)

4. Compute $f_n(k\Delta) = i^n 0.25 \sum_{m\geq 2|\mu|} J_n(k\Delta\rho_m \tilde{f}_n(\rho_m))(\rho_{m+1}-\rho_m)$, here $\rho_m \sqrt{(0.5m)^2-\mu^2}$. (7.15)

Notice that step (7.14) does not necessarily need the Bessel functions through the FFT and step (7.15) evaluates the numerical integral on an unevenly sampled grid. Thus the interpolation is taken care during the numerical integral. Detailed numerical integration methods are referred to [76].

The implementation of (3.21) is rather simple. We perform the FSE on (7.7) as follows

$$\int_0^{2\pi} f(r,\varphi)e^{-in\varphi}d\varphi$$
$$= \frac{1}{2\pi}\int_{-\infty}^{\infty}dl\int_0^{2\pi}p(\theta,l)d\theta[\int_0^{2\pi}e^{-in\varphi+\mu r\sin(\theta-\varphi)}R_\mu^{\hat{\omega}}(r\cos(\theta-\varphi)-l)d\varphi]$$
$$= \frac{1}{2\pi}\int_{-\infty}^{\infty}dl\int_0^{2\pi}e^{-in\theta}p(\theta,l)d\theta[\int_0^{2\pi}e^{-in\varphi-\mu r\sin\varphi}R_\mu^{\hat{\omega}}(r\cos\varphi-l)d\varphi].$$ (7.16)

Define kernel function $R_n(r,l)$ as

$$R_n(r,l) = \frac{1}{2\pi}\int_0^{2\pi}e^{-in\varphi-\mu r\sin\varphi}R_\mu^{\hat{\omega}}(r\cos\varphi-l)d\varphi .$$ (7.17)

We have

$$f_n(r) = \int_{-\infty}^{\infty}R_n(r,l)p_n(l)dl .$$ (7.18)

Notice that kernel function $R_n(r,l)$ can be pre-computed using the FFT, thus we assume that $R_n(r,l)$ has been available. Summarizing above procedure, the implementation of (3.21) can be carried out by those steps:

1. Compute $p_n(l)$ from $p(\theta,l)$ by the FFT. (7.19)

2. Calculate $f_n(r)$ through (7.18). (7.20)

3. Compute $f(r,\varphi)$ from $f_n(r)$ by IFFT. (7.21)

### D. Simulation results from (3.19)

The backprojection (3.24) is rather straightforward. The key step in the DBH method is to reconstruct $f(x,y)$ from $\hat{f}(x,y)$ on each vertical line. The inversion formula (2.11) may not be suitable to numerical realization. Here we introduce the numerical method form [41]. Rewrite (2.9) with $q=1$ as

$$H_\mu(s) = \frac{1}{\pi}\int_{-1}^{1}\frac{h(t)}{s-t}dt + \frac{1}{\pi}\int_{-1}^{1}A_\mu(s-t)h(t)dt, \qquad (7.22)$$

where

$$A_\mu(t) = (\cosh(\mu t)-1)/t. \qquad (7.23)$$

Apply the conventional inverse finite Hilbert transform to (7.22), we derive

$$\int_{-1}^{1}\frac{H_\mu(s)}{\pi(s-t)}\sqrt{\frac{1-t^2}{1-s^2}}ds = h(t) + \int_{-1}^{1}\Psi_\mu(t,p)h(p)dp, \qquad (7.24)$$

where

$$\Psi_\mu(t,p) = \int_{-1}^{1}\frac{A_\mu(s-p)}{\pi^2(s-t)}\sqrt{\frac{1-t^2}{1-s^2}}ds = \int_{-1}^{1}\frac{A_\mu(s-p)-A_\mu(t-p)}{\pi^2(s-t)}\sqrt{\frac{1-t^2}{1-s^2}}ds \qquad (7.25)$$

Notice $[A_\mu(s-p)-A_\mu(t-p)]/(s-t)$ is a smooth function, thus kernel $\Psi_\mu(t,p)$ defines a compact integral operator in $L^2_w(I_1)$, denoted by $\Psi$. We rewrite (7.24) in the following expression

$$[(\mathbf{I}+\Psi)h](t) = \int_{-1}^{1}\frac{H_\mu(s)}{\pi(s-t)}\sqrt{\frac{1-t^2}{1-s^2}}ds. \qquad (7.25)$$

Equation (7.25) actually constructs a Fredholm operator of the second kind. Assume $\hat{f}(x,y)$ is available in $(-1,1)\times(-1,1)$. Combining (3.21) and (7.25), we derive the integral equation

$$[(\mathbf{I}+\Psi)f](x,y) = \int_{-1}^{1}\frac{\hat{f}(\tau,y)}{\pi(\tau-x)}\sqrt{\frac{1-x^2}{1-\tau^2}}d\tau. \qquad (7.26)$$

Equation (7.26) can be solved by the simple matrix inversion as discussed in [74, 76] for the Fredholm integral equation of the second kind. Let $x_n = (n+0.5)\Delta$ be the sampling points in $[-1,1]$, here $\Delta = 1/N$ and $-N \le n < N$ with $N = 256$, using the trapezoidal rule, we discretize the integral operator $\Psi$ in the left hand side of (7.26) as

$$\int_{-1}^{1}\Psi_\mu(x_m,p)f(p,y)dp = \Delta\sum_{n=-N}^{n=N-1}\Psi_\mu(x_m,x_n)f(x_n,y). \qquad (7.27)$$

We define two vectors

$$\vec{F}(y) = \{f(x_m,y)\} \text{ and } \vec{Q}(y) = \left\{\sum_{n=-N}^{N-1}\frac{\Delta\sqrt{1-x_m^2}(x_n-x_m)\hat{f}(x_n,y)}{\pi[(x_n-x_m)^2+\Delta^2/64]\sqrt{1-x_n^2}}\right\}. \qquad (7.28)$$

Denote $\delta_{m,n}$ as the Kronecker delta symbol, and define matrix $\mathbf{M}_\mu$ as

$$\mathbf{M}_\mu = \{\delta_{m,n} + \Delta\Psi_\mu(x_m,x_n)\}. \qquad (7.29)$$

With the preceding notion, we have $\mathbf{M}_\mu \vec{F}(y) = \vec{Q}(y)$. Then $\vec{F}(y)$ can be solved by the equation systems

$$\vec{F}(y) = \mathbf{M}_\mu^{-1}\vec{Q}(y). \qquad (7.30)$$

Equation system (7.30) must be familiar to readers with little mathematics or engineering background. In our numerical studies, matrix $\mathbf{M}_\mu^{-1}$ seems to be well defined, thus the simple matrix inversion is sufficient to solve (7.26). The intermediate image $\hat{f}(x,y)$ and the reconstructed image $f(x,y)$ are shown in Fig 8. The reconstructed image by the DBH method from half-scan data is comparable to the reconstructed image by the FBP algorithm from full-scan data.

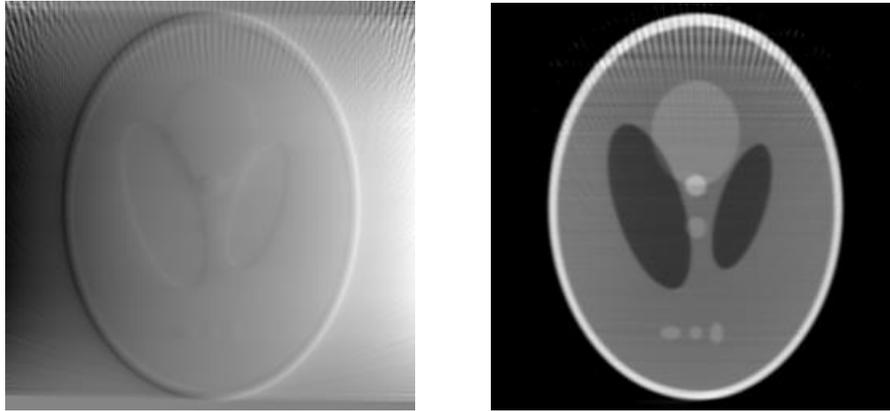

Fig 8: Left is the intermediate image $\hat{f}(x,y)$ and right is the reconstructed image using (7.30).

### E. Projection noise estimate and effect of reconstruction algorithms

So far, we have assumed the projection data to be the exponentially weighted line integrals. The actual data from the physical instruments never become that perfect. For example, due to the Poisson nature of the Gamma photon emissions, the actual measurements are integer counts instead of floating numbers. In the discrete case, at each line of $(\theta_k, s_n)$, we assume that the measurement $\tilde{p}(\theta_k, s_n)$ follows the Poisson law with mean value of $p(\theta_k, s_n)$. Then the standard deviation of $\tilde{p}(\theta_k, s_n)$ is $1/\sqrt{p(\theta_k, s_n)}$. This indicates that the difference between $\tilde{p}(\theta_k, s_n)$ and $p(\theta_k, s_n)$ could become severe if $p(\theta_k, s_n)$ is small. In our numerical simulation study, we assume that $p(\theta_k, s_n)$ is mathematically exact, and measurements $\tilde{p}(\theta_k, s_n)$ are from the Poisson counts generator in [76]. Readers should remember that the independency assumption is used in the Poisson counts generator. Two sets of measurements $\tilde{p}(\theta_k, s_n)$ with different levels of noise are shown in Figure 9.

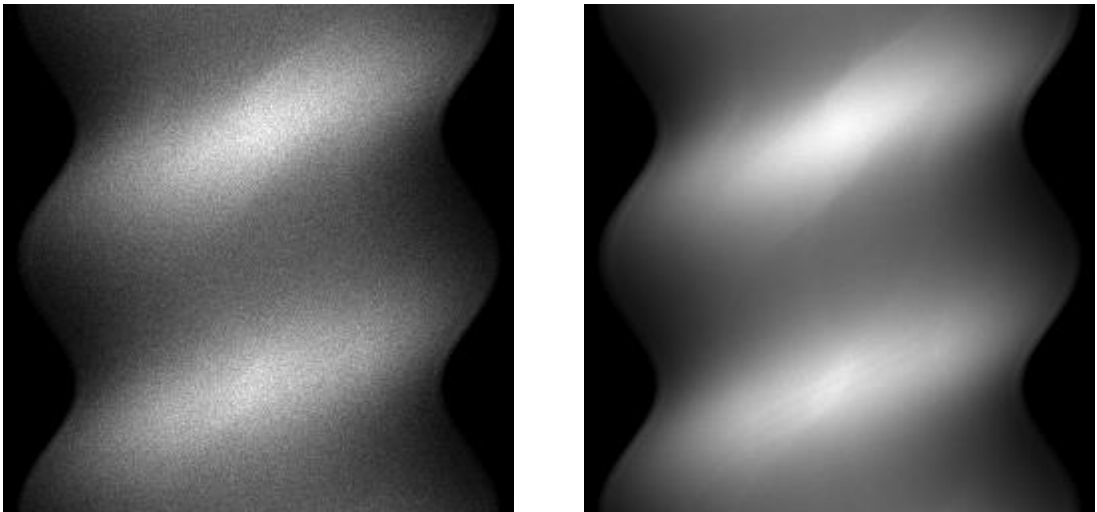

Fig 9: Left is the Poisson counts data and right is the median-filtered data.

Now we use two sets of noisy projection data to verify the amazed accuracy from (6.3). The estimated *err* by using (6.3) are 6.55% and 1.23%, respectively. The actual *err* of these two sets of projection data are 6.56% and 1.22%, respectively. It is quite amazing about such high accuracy. Combining relations (6.2) and (6.3), one derives the following approximation

$$\frac{\sum_{k,m} n^2(\theta_k, s_m)}{\sum_{k,m} p^2(\theta_k, s_m)} \approx \frac{\sum_{k,m} \tilde{p}(\theta_k, s_m)}{\sum_{k,m} \tilde{p}^2(\theta_k, s_m) - \sum_{k,m} \tilde{p}(\theta_k, s_m)}. \tag{7.31}$$

The authors of [58] further explore (7.31) for designing digital filters to smooth the noise. Denote by **F** a digital filter. The criterion used in [58] is to find a specific form of **F** satisfying

$$\frac{\sum_{k,m}[\tilde{p} - \mathbf{F}\tilde{p}]^2(\theta_k, s_m)}{\sum_{k,m}[\mathbf{F}\tilde{p}]^2(\theta_k, s_m)} \approx \frac{\sum_{k,m} \tilde{p}(\theta_k, s_m)}{\sum_{k,m} \tilde{p}^2(\theta_k, s_m) - \sum_{k,m} \tilde{p}(\theta_k, s_m)}. \tag{7.32}$$

One question in using criterion (7.32) is that the option of selecting **F** is too arbitrary without restricting the function class that **F** falls in. The sinc-like kernel functions was evaluated in [58]. In our numerical study, it was confirmed that the median filter is another method that closely meets (7.32). Regardless of the class of **F**, the criterion (7.32) lacks a close connection to the ultimate goal in (6.7) to reduce the noise in the reconstructed image. The recent work [60] does not guarantee the reduction of the noise in the reconstructed image either. In author's view, relation (6.3) is a very useful tool to estimate the noise level only based on the measurements, but may be limited in designing the filters.

Equation (6.5) indicates that the total noise $n(\theta_k, s_m)$ can be decomposed into range space noise $n_R(\theta_k, s_m)$ and null space noise $n_N(\theta_k, s_m)$. The null space noise can be estimated by (6.6). This indeed provides insights to better understand the noise. Since simulated measurements $\tilde{p}(\theta_k, s_n)$ are based on exact projection $p(\theta_k, s_m)$, thus both $p(\theta_k, s_m)$ and $\tilde{p}(\theta_k, s_m)$ are known. The numerical results indicate that $err(n_N, n) = 0.38$ and $err(n_R, n) = 1.02$. This implies that the null space noise is the major component. This observation contradicts the common wisdom that reducing the projection noise should help suppress the noise in the reconstructed image. More simulation results and analysis of the range condition can be found in [59]. The range condition may shed the light of the complexity in the noise propagation from projection to the reconstructed image, but does not help find a concrete solution to filter the noise.

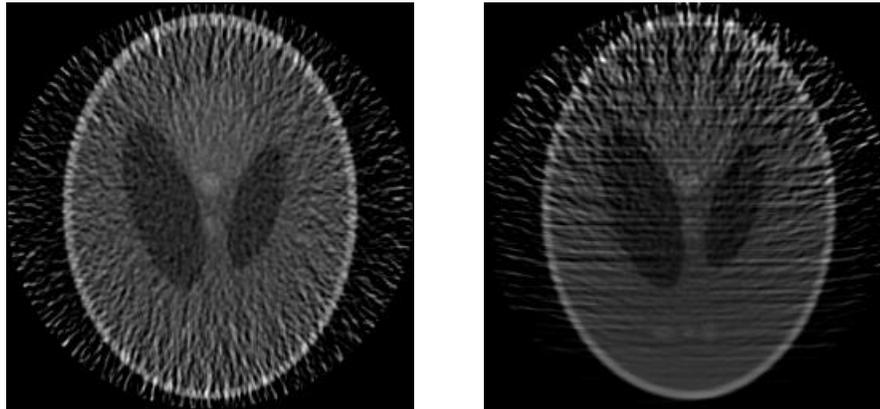

Fig 10: Left is the reconstructed image by (7.7) and right is the reconstructed image using (7.30).

So far, there has been no mathematical theory to guarantee that reducing projection noise would mitigate the noise in the reconstructed image. Nonetheless, different reconstruction algorithms do reveal quite different numerical characteristics. For example, the reconstructed images from the FBP and DBH algorithms show quite different noise pattern as shown in Fig 10, in which three small ellipses are barely recognizable in the image from the FBP method but remain well observable in the image by the DBH method. Certainly, such difference could make immediate effect to improve the lesion detectability in the clinic SPECT study.

## VIII. Discussions and Future Works

The central slice theorem used to be the key relation for developing analytical reconstruction algorithms, and has been well investigated in the literature. Recently, three new techniques of wave propagation equation, 2D fundamental solution and finite Hilbert transform have been developed to handle the attenuation correction and the half-scan data reconstruction. The DBH method translates the 2D image reconstruction into the inversion of a finite weighted Hilbert transform through the DBP operation. More developments on this method have been carried out actively in two areas: SPECT [38-42] and CT [82-83]. Based on [35], the extension of the DBH method to MRI image reconstruction seems to be another interesting topic. At the same time, the numerical stability and noise effect of the DBH method have not been thoroughly studied except some observations. This may be one area worth further investigations from an application point of view. Again, for the half-scan data issue, exploring the existence of a 2D finite inversion of (3.31) is another important open issue from theoretical interests and medical applications of handling non-parallel projections.

Katsvich derived a 1D filtering procedure for the exact reconstruction of helical cone-beam data in [80, 81], and soon triggered a wave of research on the exact reconstruction of fan- and cone-beam data, see the brief review in [84] and the references therein. The combination of Katsvich's idea and the DBH method seems very likely to yield an exact reconstruction of the helical cone-beam ERT. Some initial results were reported in [91], but more efforts are needed to have a definitive answer to whether the DBH method is applicable to the helical cone-beam ERT.

The SVD method for the RT in [85] is very beautiful. The recent work [92] provides the series expansion of the attenuated Radon transform by using Zernike polynomials. The author expects to have the similar series expansion for the ERT with limited-angle data.